\documentclass[journal=jacsat,manuscript=article]{achemso}
\setkeys{acs}{etalmode=truncate,maxauthors=10}
\mciteErrorOnUnknownfalse

\usepackage[version=3]{mhchem} 
\usepackage{amssymb}
\usepackage{xcolor}
\usepackage[colorlinks=true,breaklinks=true,linkcolor=black,citecolor=green,urlcolor=magenta]{hyperref}
\usepackage{siunitx}
\usepackage{graphicx}

\usepackage{xcolor}
\usepackage[capitalise]{cleveref}
\usepackage{algorithmicx}
\usepackage{algorithm}
\usepackage[noend]{algpseudocode}
\usepackage[normalem]{ulem}
\usepackage{makecell}

\algnewcommand\algorithmicinput{\textbf{Input:}}
\algnewcommand\Input{\item[\algorithmicinput]}

\algnewcommand\algorithmicoutput{\textbf{Output:}}
\algnewcommand\Output{\item[\algorithmicoutput]}

\usepackage{bm}
\usepackage{braket}

\newcommand{\vett}[1]{{\bf{#1}}}
\newcommand{\bts}{\vett{x}}

\newcommand{\nocontentsline}[3]{}
\newcommand{\tocless}[2]{\bgroup\let\addcontentsline=\nocontentsline#1{#2}\egroup}

\title{Quantum-Centric Alchemical Free Energy Calculations}
\author{Milana Bazayeva}
\affiliation{Center for Computational Life Sciences, Lerner Research Institute, The Cleveland Clinic, Cleveland, Ohio 44106, United States}
\author{Zhen Li}
\affiliation{Center for Computational Life Sciences, Lerner Research Institute, The Cleveland Clinic, Cleveland, Ohio 44106, United States}
\author{Danil Kaliakin}
\affiliation{Center for Computational Life Sciences, Lerner Research Institute, The Cleveland Clinic, Cleveland, Ohio 44106, United States}
\author{Fangchun Liang}
\affiliation{Center for Computational Life Sciences, Lerner Research Institute, The Cleveland Clinic, Cleveland, Ohio 44106, United States}
\author{Akhil Shajan}
\affiliation{Center for Computational Life Sciences, Lerner Research Institute, The Cleveland Clinic, Cleveland, Ohio 44106, United States}
\affiliation{Department of Chemistry, Michigan State University, East Lansing, Michigan 48824, United States}
\author{Susanta Das}
\affiliation{Center for Computational Life Sciences, Lerner Research Institute, The Cleveland Clinic, Cleveland, Ohio 44106, United States}
\author{Kenneth M. Merz Jr.}
\email{kmerz1@gmail.com}
\affiliation{Center for Computational Life Sciences, Lerner Research Institute, The Cleveland Clinic, Cleveland, Ohio 44106, United States}
\affiliation{Department of Chemistry, Michigan State University, East Lansing, Michigan 48824, United States}

\begin{document}
\begin{center}
    \textit{M.B. and Z.L. contributed equally to this work.}
\end{center}

In the present work, we present a hybrid quantum-classical workflow aimed at improving the accuracy of alchemical free energy (AFE) predictions by incorporating configuration interaction (CI) simulations using the book-ending correction method. This approach applies the Multistate Bennett Acceptance Ratio (MBAR) over a coupling parameter $\lambda$ to smoothly transition the system from molecular mechanics (MM) ($\lambda$ = 0) to a quantum mechanics (QM) ($\lambda$ = 1) description. The resulting correction is then applied to the classically (MM) computed AFE to account for the more accurate QM treatment. The standard book-ending procedure uses AMBER to simulate the MM region, and QUICK, AMBER's default QM engine, to handle the QM region with either the Hartree-Fock (HF) method or density functional theory (DFT). In this work, we introduce a novel interface to QUICK, via sander, that enables CI simulations, and can operate in two ways: A) via PySCF backend to perform full configuration interaction (FCI) using conventional computing resources, B) quantum-centric sample-based quantum diagonalization (SQD) workflow via Qiskit which leverages both quantum hardware and post-processing on conventional computing resources for CI simulations. In this workflow QUICK performs most steps of the calculations, but at user-defined intervals, it redirects the computation to either FCI or SQD backend to get the CI result. We computed the book-end corrections for the hydration free energy (HFE) of three small organic molecules (ammonia, methane, and water) to benchmark the proposed approach and demonstrate how quantum-computers can be used in AFE calculations. We believe that this approach can be scaled to more complex systems like drug-receptor interactions in future studies.


\section{Introduction}

Accurate prediction of molecular binding affinities is fundamental in drug discovery. Among all the available computational strategies, alchemical free energy (AFE) calculations \cite{afe} stands out as a powerful method to estimate key molecular properties, such as  hydration free energies (HFE) and ligand-receptor binding affinities~\cite{devivo2016, jorgensen2000, li2024, saloahen2021}. 
However, the accuracy of current AFE approaches remains limited by the approximations inherent in classical computational methods, particularly the selection and parametrization of force fields (FFs). This is becoming more evident as drug discovery extends into more complex environments involving ions, nucleotides, and interacting molecules \cite{zeng2023, giese2022}.
Moreover, classical approaches do not include subtle electronic and polarization effects occurring in such complex systems \cite{clemente2023, Kar2023}.  

Recent advancements in quantum chemistry have significantly impacted drug discovery and free energy calculations, offering promising avenues for accelerating research and improving accuracy using the QM/MM interface~\cite{warshel1976, duarte2015, Kar2023, manathunga2022, Lu01092016}.
However,  direct AFE calculations using QM/MM methods remain challenging due to the computational demand of the QM calculations arising from the need of extensive configurational sampling \cite{giese2024, york2023,Li2003}. In this context, the book-ending approach has emerged as an effective strategy to overcome the aforementioned limitations of the fully QM/MM AFE calculations. The core of this indirect approach is to compute the AFE leveraging classical MM force fields and existing MM-based free energy approaches. To obtain QM/MM accuracy, a correction of the two end-states is applied. For each state, the free energy difference between the MM and QM/MM description is computed and combined with the classical result. The book-ending correction requires less sampling since the chemical composition is not affected during the transformation. \cite{giese2024} 

Building on the work  of Gao \cite{Gao1992,gao1992priori} and Warshel \cite{Warshel1992} on the use of reference potentials, and further developed by Woodcock, Boresch, König, Brooks. \cite{molecules24040681,hudson2018force,hudson2018accelerating,Kearns2017,konig2014,konig2014multiscale,konig2015,konig2018,konig2018comparison,hudson2015}, Giese et al. proposed the book-ending approach for force field parametrization. \cite{giese2019} Here, the MM force field parameters are optimized to match the forces computed at QM/MM level of theory through a series of reference potentials. Recently, a machine learning (ML) workflow was introduced in book-ending to further improve the accuracy.\cite{giese2024} However, the predicted free energies still deviate from the experimental data. This highlights the importance of employing more accurate ab initio methods, as explored in the present work. 

While density functional theory (DFT) is often chosen as a reference method in such workflows for the good trade-off between computational cost and performance, it can fail in describing systems with high accuracy. \cite{dftrew} In contrast, the FCI method provides the exact solution of the CI problem within the chosen basis set and contains all the possible electron configurations. For this reason, FCI is used as a standard in the assessment of less accurate quantum methodologies.\cite{Gao2024, Vogiatzis2017} FCI is computationally prohibitive thus, is infeasible for large systems. This makes heuristic and approximate methods necessary to obtain results within a practical computational time. One such method is selected CI (SCI) approach, which retains only the most energetically significant determinants to reach the near-FCI level of accuracy with a much more favorable computational expense.\cite{Abraham, Gao2024} However, the classical heuristics used in SCI to select the determinants, can be inefficient during the subspace definition, leading to reduced accuracy and increased computational overhead. \cite{sharma2017semistochastic} Both FCI and SCI being NP-hard problems on classical hardware, push toward the exploration of alternative strategies for more efficient exploration of electronic configuration space.\cite{holmes2016heat} 

In the proposed approach, we apply the quantum-centric SQD methodology to AFE calculations by integrating it within the book-ending workflow. Sample-Based Diagonalization (SQD) \cite{robledo2024chemistry,shajan2024towards,Liepuoniute2024,Barison2025,yu2025quantum} represents a promising strategy to overcome the limitations of both FCI and SCI, offering scaling opportunities and near-FCI accuracy even for large system that are still inaccessible to classical approaches. To enable efficient coupling between quantum processing unit (QPU) and classical CPUs, we developed a dedicated interface. The same interface also supports conventional FCI simulations, which can be used to benchmark quantum-centric simulations. Our FCI- and SQD-based book-ending framework can be further scaled up with density matrix embedding theory (DMET), the efficiency of combining SQD and DMET was demonstrated in our recent study. \cite{shajan2024towards} Additionally, the successful application of SQD to solvated systems under implicit solvent conditions \cite{kaliakin2025implicit} supports its efficiency and motivates its extension in the more challenging context of AFE calculations. Our interface combines the strengths of Fortran90 and Python 3.11. FORTRAN enables high-performance calculations in AMBER and the ab initio electronic structure program QUICK, while Python grants a user-friendly interface for setting up and running quantum-centric SQD simulations on real quantum hardware. We note that the present paper is dedicated to establishing the CI-corrected book-ending framework. The systems considered in the present paper do not include the strong electron correlation effects. \cite{Raghavachari,Martin} Hence, this work represents the baseline benchmark which will pave the way for CI-based book-ending calculations in more complex systems.

\section{Methods and Computational Details}

\subsection{Structure preparation and MM HFE calculation}
\label{MMHFE}

The initial coordinates for ammonia, methane, and water were generated using the LEaP module of the AMBER24 software package~\cite{amber24}. Ammonia and methane were parameterized using the General AMBER Force Field (GAFF)~\cite{wang2020gaff2}, with atom types and bonded parameters automatically assigned by LEaP. Atomic partial charges for these molecules were derived using the Restrained Electrostatic Potential (RESP) fitting method~\cite{woods2000resp}, based on molecular electrostatic potentials calculated at the B3LYP level with the 6-31G* basis set using the Gaussian software package \cite{g16}. The water system was set up using the OPC3 \cite{onufriev2016opc3} parameters included in AMBER. The final system of each solute included the corresponding dummy molecule used to enable the alchemical transformation for the free energy calculations. The final system was then embedded in a cubic box filled with water molecules, ensuring a minimum 24 Å padding from the solute in each direction. Periodic boundary conditions (PBC) were applied to eliminate edge effects and mimic bulk solvent environment. The OPC water model \cite{onufriev2014opc} was used for ammonia and methane, while the OPC3 \cite{onufriev2016opc3} one was applied to the water system. Every system underwent a two-step energy minimization: an initial steepest descent phase followed by conjugate gradient minimization, both consisting of 10,000 steps. The Electrostatic contribution was treated using the Particle Mesh Ewald (PME) method~\cite{darden1993pme} with a real-space cutoff of 10 Å. The FFT grid used for PME was set to 48 × 48 × 48 points.  An NVT equilibration of 360 ps was then performed, during which each system was gradually heated from 0 K to 300 K in 50 K increments. Temperature control was achieved using Langevin dynamics with a collision frequency of 2 ps-1, providing stable and efficient thermal regulation during the heating phase. Following temperature equilibration, an NPT equilibration at 300 K and 1 atm was carried out for 300 ps to further relax the system. Pressure was regulated using the Berendsen barostat \cite{berendsen1984bath} with a reference pressure of 1 bar. The thermodynamic integration was activated from the minimization stage through all equilibration and production stages by applying the specific mask for the real molecule and its corresponding dummy. The fully equilibrated structures were used in the subsequent classical HFE calculation and book-ending correction. The former was obtained by performing a even-window Thermodynamic Integration (TI) for the three molecules of interest (ammonia, methane, and water), following the equation below:

\begin{equation}\label{eq1}
    U(\lambda) = (1-\lambda)U_0 + \lambda U_1
    \;.
\end{equation}

In Eq.~\eqref{eq1}, $U(\lambda)$, $U_0$, and $U_1$ denote the overall energy of the perturbed system at a specific $\lambda$ window, the energy of the initial state ($\lambda=0$, which is the molecule dissolved in water), and the energy of the final state ($\lambda=1$, which is the molecule "disappeared" to a dummy molecule in water). The $\Delta A$ between the initial and final system can then be expressed as: 
\begin{equation}\label{eq2}
    \Delta A = \int_{0}^{1} \left\langle \frac{dU}{d\lambda} \right\rangle \,d\lambda
\end{equation}

Because of the ensemble energy transition from the initial state to the final state follows Gaussian quadrature rules. We can rewrite Eq.~\eqref{eq2} to
\begin{equation}\label{eq3}
    \Delta A = \sum_{i=1}^{7} \left( c_i \times \left\langle \frac{dU_i}{d\lambda_i} \right\rangle \right)
\end{equation}

In Eq.~\eqref{eq3}, $c_i$ are the Gaussian quadrature coefficients \{0.065, 0.140, 0.191, 0.209, 0.191, 0.140, 0.065\} and $\lambda_i$ are the discrete values \{0.025, 0.130, 0.297, 0.500, 0.703, 0.870, 0.975\} to mix the initial and final values. The whole TI simulation takes 21 ns (3 ns for each of the 7 $\lambda$ windows) under NVT to obtain the Helmholtz free energy, then converted to Gibbs free energy as in the NPT ensemble. 

\subsection{Book-ending simulations}

\begin{figure}
     \centering
     \includegraphics[width=385pt]{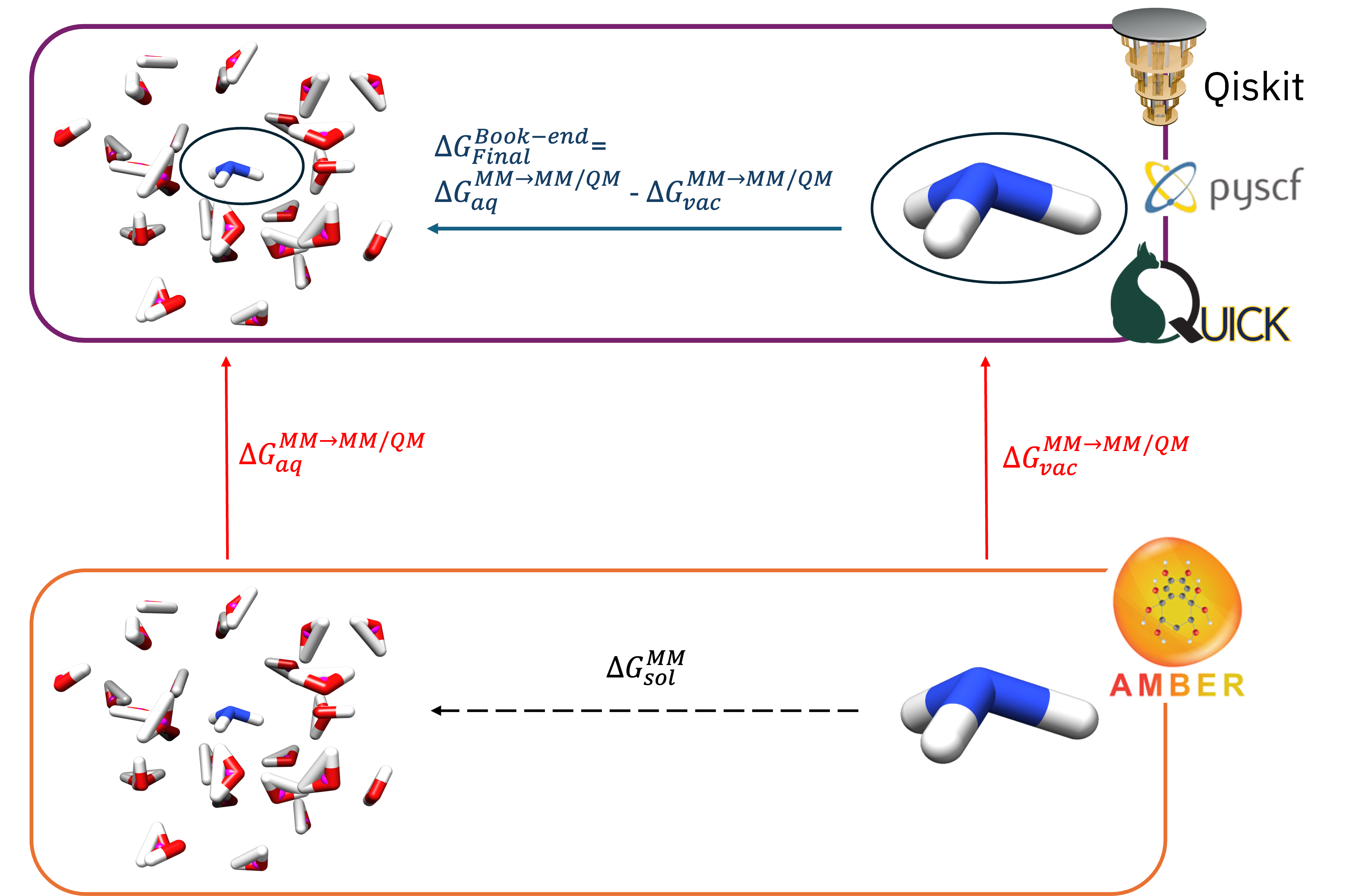}
     \caption{Schematic representation of the book-ending energy correction procedure. At the bottom, the HFE is computed using classical methods (dashed black arrow). The book-ending approach employs thermodynamic integration (TI) to switch the potential from MM to QM/MM at each end-state (vertical red arrows). The quantum contribution can be calculated using one of three approaches: HF in QUICK package, FCI in PySCF package, or the quantum-centric SQD approach through Qiskit Addon: SQD. The free energy difference obtained from TI is analyzed with MBAR to yield the quantum correction, which is then added on top of the classical HFE (blue arrow).}
     \label{fig1}
\end{figure}

As mentioned previously, the book-ending correction method is an indirect approach to achieve QM accuracy in AFE calculations without performing fully QM/MM simulations throughout the entire alchemical pathway. Instead, it exploits MM simulations combined with a correction computed only at the end-states of the transformation (Figure~\ref{fig1}). 
This correction is evaluated by gradually transforming the system from a purely MM potential  (${\lambda}$=0) into a QM/MM description  (${\lambda}$=1) via intermediate states. The MBAR analysis is used to compute the free energy difference between the two levels of theory and provides a correction to refine the classically computed HFE values. Because only the potential energy function changes and not the atomic configurations, the method achieves high accuracy with minimal sampling effort. Here, we prepared the initial coordinates for ammonia, methane, and water following the same protocol described previously for the classical HFE calculations. However, no dummy molecules and no thermodynamic integration settings were used during the preparation of the starting structures. Hydrogen mass repartitioning (HMR) \cite{hmr} was applied to allow 1 fs as the integration timestep. For each equilibrated solute (ammonia, methane, and water) six ${\lambda}$ windows (0.00, 0.20, 0.40, 0.60, 0.80, 1.00) were simulated in a NVT ensemble using PBC.  The Langevin dynamics with a collision frequency of 5 ps was used to maintain the target temperature of 298 K. No pressure control was applied, and a nonbonded interaction cutoff of 10 Å was used. The SHAKE algorithm was applied to constrain bonds involving hydrogen, with the exception of the solute, which was described quantum mechanically. The book-ending simulations use two separate input files per ${\lambda}$ window, which are identical in all respects except that one performs a classical MM calculation, while the other enables QM/MM calculations with electrostatic embedding. The QM region, which included the solute molecule, was described at the Restricted Hartree-Fock (RHF) level with the STO-3G basis set, using QUICK engine \cite{manathunga2023quantum,cruzeiro2021open} for the QM treatment throughout all simulations. At specified interval, an external CI solver can be invoked to evaluate the QM forces. This enables the use of either  FCI methods or quantum-centric calculations to achieve near-FCI accuracy (Figure~\ref{fig2}). Further details are provided in the  Methods section below and in the Supplementary Information. Each ${\lambda}$ window was equilibrated for 1 ps and was followed by 1 ps production run. During equilibration, each window was dependent on the preceding one. On the other hand, the production runs were started independently for each ${\lambda}$ window, using only their corresponding equilibrated structure. 
Each solute was simulated for a total of  6 ps across the ${\lambda}$ windows (1 ps per window). Simulations were performed under three different computational setups: A) with QUICK alone, B) QUICK coupled to an FCI solver, C) QUICK interfaced with quantum hardware and classical post-processing via the SQD method. For each setup, three independent replicates were performed for statistical analysis. The external CI solvers were activated only during the production runs, at every 10\textsuperscript{th} step.

\subsection{Multistate Bennett Acceptance Ratio (MBAR) analysis}
The MBAR calculations \cite{mbar} are often derived from 
\begin{equation}\label{eq4}
    \Delta A_{ij} = \frac{\ln\frac{Q_j}{Q_i}}{-\beta}
    \;.
\end{equation}

In Eq.~\eqref{eq4}, $\Delta A_{ij}$, denotes the relative Helmholtz free energy difference between two designated states, $\lambda_i$, $\lambda_j$ of the multistate system, where $Q_i$ and $Q_j$ denote the partition functions of the two states, as expressed in the Poisson-Boltzmann distribution. Here, $\beta$ is defined as $1/(k_BT)$ , where $k_B$ is the Poisson-Boltzmann constant and $T$ is the absolute temperature. Re-writing Eq.~\eqref{eq4}, we obtain
\begin{equation}\label{eq5}
    Q_i\langle\alpha_{ij}\times e^{-\beta U_j}\rangle_i = Q_j\langle\alpha_{ij}\times e^{-\beta U_i}\rangle_j 
\end{equation}

In Eq.~\eqref{eq5}, $\alpha_{ij}$ is the probability that a sample being drawn from state $\lambda_j$ appears to follow the distribution of partition functions from state $\lambda_i$. A larger $\alpha_{ij}$ value usually means a  better overlap between states $\lambda_i$ and $\lambda_j$. When sampling is sufficient, according to the definition of $\langle\alpha_{ij}\times e^{-\beta U_j}\rangle_i$, $\alpha_{ij}$ can be further expressed as below in an $i$-independent way

\begin{equation}\label{eq6}
    \alpha_{ij} = \frac{\frac{N_j}{\hat{c}_j}}{\sum\limits_{k=1}^K \frac{N_k  e^{\left(-\beta U_k \right)}}{\hat{c}_k}}
\end{equation}

In Eq.~\eqref{eq6}, $K$ is the total number of $\lambda_k$ states (in this work, we selected six, which is from $\lambda_1=0.0$, $\lambda_2=0.2$ ... to $\lambda_6=1.0$). $N_k$ denotes the total number of samples drawn at state $\lambda_k$, and $\hat{c_k}$ denotes the partition function of state 
$\lambda_k$ up to a constant (which is the arbitrarily defined acceptance ratio in MBAR).

Inserting Eq.~\eqref{eq6} back to Eq.~\eqref{eq5} and finally to Eq.~\eqref{eq4}, we have the self consistency expression of 

\begin{equation}\label{eq7}
     \hat{A_i} = \frac{\ln \sum\limits_{j=1}^K \sum\limits_{i=1}^{N_j} \frac{e^{-\beta U_i}}{\sum_{k=1}^K N_k e^{\beta\hat{A_k} - \beta U_k}}}{-\beta}
\end{equation}

In Eq.~\eqref{eq7}, once we have an initial guess of $\{\hat{A_k}, k\in \{1,2,3,...,6\}\})$ since we have six $\lambda$ values, we will then be able to iteratively update each $\hat{A_k}$ using Eq.~\eqref{eq4} until the optimized Helmholtz free energy is obtained in the NVT ensemble, then converted to the Gibbs free energy in the NPT ensemble. The whole calculation and analysis is automatized via the AMBER24 package~\cite{amber24}. The final corrected bookending energy $\Delta \Delta G_{bookending}^{QM/MM}$ is then added on top of the MM HFE ~\ref{MMHFE}as Figure~\ref{fig1} shows.

\subsection{Interfacing AMBER with external CI solvers}

\begin{figure}
    \centering
    \includegraphics[width=1\linewidth]{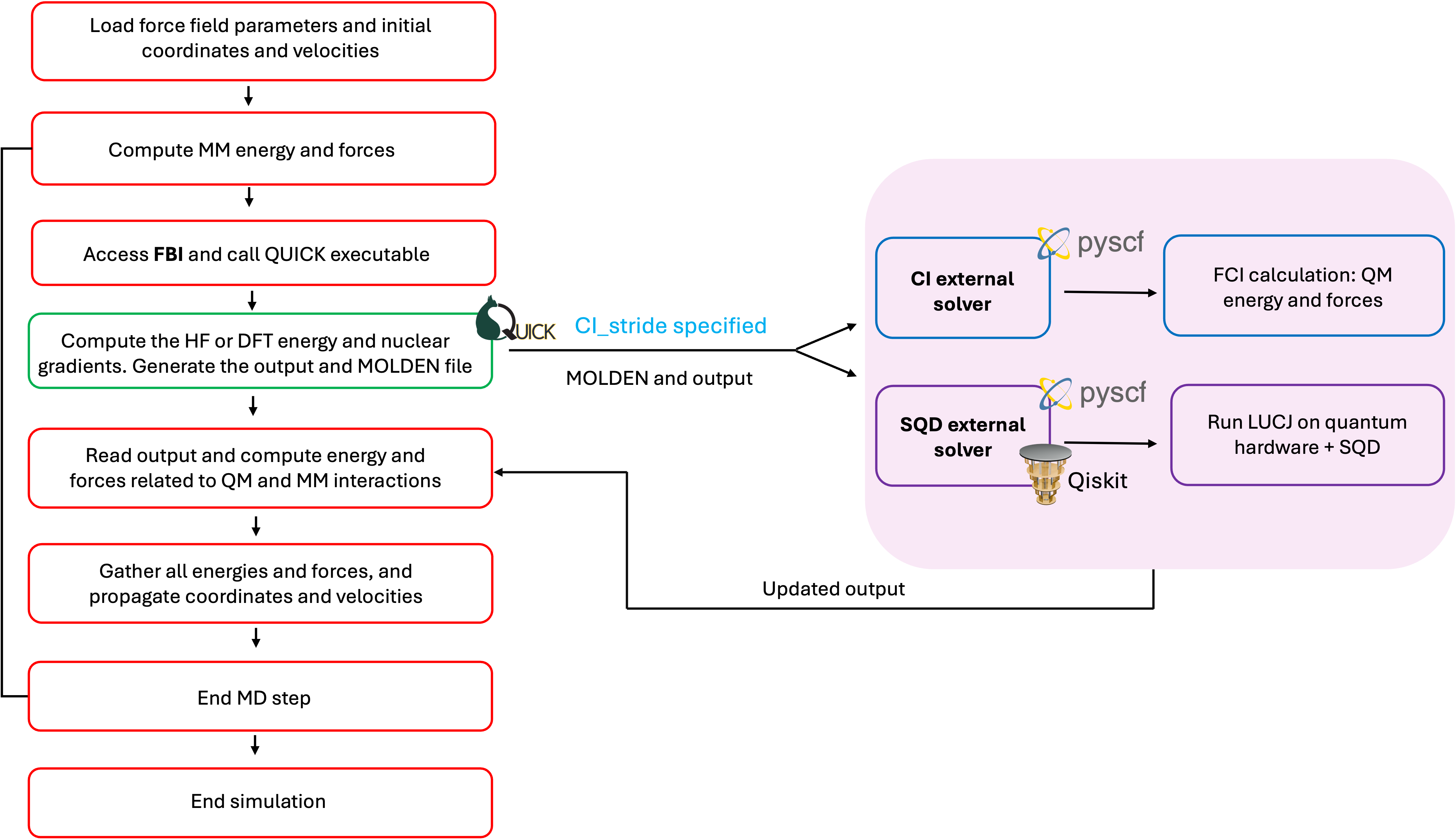}
    \caption{
    Workflow of the standard sander functionality within the QM/MM scheme (left), extended with the interface proposed in this work (right). 
    The steps managed by sander are shown in red, those performed by QUICK engine in green, and those handled by external CI solvers are enclosed in the purple box. The FCI simulations are performed using PySCF (boxes in light blue), while for quantum-centric SQD calculation leverage a synergetic combination of PySCF and Qiskit (boxes in purple).}
    \label{fig2}
\end{figure}

To enable coupling between classical molecular dynamics (MD) and quantum circuit-based computations, we designed a modular extension of the AMBER/QUICK (via sander) framework that allows for the treatment of the QM region using either FCI solver or via quantum-centric SQD simulations. We developed a custom interface that connects AMBER/QUICK with the Qiskit ecosystem \cite{aleksandrowicz2019qiskit,javadi2024quantum}, facilitating access to quantum hardware and subsequent SQD post-processing.\cite{robledo2024chemistry, kaliakin2024accurate,yu2025quantum,Liepuoniute2024,Barison2025,barroca2025surface,yu2025quantum} Moreover, we leveraged the same interface to introduce conventional FCI calculations, which we used to benchmark of our quantum-centric results (Figure~\ref{fig2}). 

Both approaches are integrated into the existing file-based interface (FBI) architecture of QUICK \cite{manathunga2023quantum,cruzeiro2021open}, which allows the evaluation of the QM energies and gradients using the HF method at each step. This approach enables a CI-based correction to be applied periodically along the trajectory. Specifically, at user-defined intervals (controlled by the CI\_stride keyword), the QM simulation is redirected towards either: A) an external FCI solver implemented through the interface with PySCF; B) the SQD solver that utilizes the quantum hardware to generate the initial electron configurations with the LUCJ Ansatz \cite{motta2023bridging} and subsequent post-processing through Qiskit Addon: SQD. \cite{robledo2024chemistry, kaliakin2024accurate,yu2025quantum,Liepuoniute2024,Barison2025,barroca2025surface,yu2025quantum} Detailed description of SQD simulations is reported in the Supplementary Information. Both options are implemented through Python scripts, where the passing of the HF orbitals from QUICK to conventional CI or SQD workflow is handled with the PySCF modules \cite{sun2018pyscf,sun2020recent}. In the standard approach (left side of Figure~\ref{fig2}), sander coordinates the classical and the quantum components. At each time step, sander computes the classical contributions of all the MM atoms. Then, the FBI is used to invoke QUICK, which evaluates the energy and nuclear gradient for the QM region using HF or DFT methods. Once QUICK has completed, sander proceeds to integrate the results into the standard output format and completes the MD computation of the step by calculating the electrostatic and van der Waals forces between the MM and the QM regions. With all the forces and energies gathered, sander propagates the atomic coordinates and velocities, advancing the trajectory. This procedure is then repeated at every MD step. On the right side of Figure~\ref{fig2}, the workflow retains the same scheme described above, in which QUICK evaluates HF energies and nuclear gradients at each MD step, while sander integrates the resulting forces with the QM/MM interactions to propagate the atomic coordinates. However, at CI\_stride defined intervals, the workflow redirects the QM computation to an external CI solver. This is done by leveraging the MOLDEN file generated by QUICK, which contains geometrical and molecular orbital information required by both external CI solvers. The energies and gradients computed by the solvers are reintegrated into the AMBER workflow, with sander continuing the MD propagation using the updated total energies and preserving the efficiency of the original QUICK interface. For more details about the development of the interface, refer to the Supplementary Information.


\section{Results and Discussion}
In the present work, we developed an interface to enable the computation of book-ending corrections using FCI, with the aim of achieving a higher level of QM accuracy. FCI provides the most accurate QM solution, but its computational cost makes it impractical for larger systems. To address this limitation, we also implemented a quantum-centric SQD approach. This alternative enables energy evaluation at near-FCI accuracy with potential scaling improvements. To validate the interface and evaluate its impact on the accuracy of book-ending corrections, we classically computed the hydration free energies (HFEs) and applied quantum-level corrections for three small molecules: ammonia, methane, and water. In all quantum-corrected protocols, equilibration was carried out using AMBER’s native QUICK engine and the HF method. In production runs, three separate protocols were applied. The basic protocol included the quantum correction with the HF method only, while in the FCI or SQD protocols, the molecular dynamics was propagated using QUICK for the majority of time steps, but FCI or SQD energy and gradients were computed and incorporated in the dynamics at every 10\textsuperscript{th} step via PySCF (in the case of FCI) and Qiskit Addon: SQD (in the case of SQD). We denote the three book-ending protocols as HF correction, HF + FCI correction, and HF + SQD correction. All QM calculations were performed with the STO-3G basis set. Table~\ref{table1} presents a comprehensive overview, including the MM-calculated HFE, the book-ending corrected values, and the Minnesota Solvation Database (MNSol) values. \cite{Manerich2020} As evident from the Table~\ref{table1}, classical MM calculations exhibit significant deviations from the MNsol database, with errors reaching up to 2.65 kcal/mol for the water system. 
The case of ammonia is the most notable. The classical HFE prediction (-3.87 kcal/mol) underestimates the MNsol database value (-4.29 kcal/mol) by approximately 0.42 kcal/mol, suggesting that a quantum correction should improve the energy. However, all quantum corrections move the result further from the reference, introducing a positive correction. Interestingly, the HF + FCI result (-2.35 kcal/mol) is the closest to the MNSol database, followed by the HF + SQD corrected value (-2.20 kcal/mol) and the HF alone (-1.94 kcal/mol). This suggests that although all methods perform consistently, none of the quantum corrections fully recover the MNSol database value. Most likely, this shortcoming is due to the minimal STO-3G basis set used in this work. In our future studies, we plan to further extend the basis set size. Finally, we note that the Lennard-Jones parameters used could be re-optimized to better fit with book-ending corrections at the level of theory used. For methane, a nonpolar and electronically simple molecule, the MM prediction (2.28 kcal/mol) is already close to MNSol database (2.00 kcal/mol). Here, the HF protocol yields a correction of -0.15 kcal/mol (Table~S2, Supplementary Information), reducing the prediction to 2.13 kcal/mol. The HF + SQD protocol adjusts the result of -0.37 kcal/mol, which slightly underestimates the MNSol reference but remains within a reasonable margin. In contrast, the HF + FCI protocol produces a slightly larger correction (-$0.84 \pm 0.33$ kcal/mol). Despite this over correction, all methods yield results within approximately 0.3–0.6 kcal/mol of the MNSol database value. For water, the classically computed HFE (-8.96 kcal/mol) overestimates the MNSol database solvation energy (-6.31 kcal/mol) by 2.65 kcal/mol. For this system all the computed book-ending corrections improve the final value. The HF protocol correction (1.94 kcal/mol) is the largest, followed by HF + SQD (1.52 kcal/mol), and HF + FCI (1.38 kcal/mol) corrections. The HF corrected value (-7.02 kcal/mol) shows the best agreement with MNSol database, with an absolute error of $\sim 0.7$ kcal/mol. Although HF + FCI protocol underestimates the required correction and gives less accurate results (-7.58 kcal/mol, within 1.30 kcal/mol from the reference value), it still improves the classically computed HFE. Moreover, the resulting correction can be further improved by increasing the number of samples in SQD method or tuning of the Lennard-Jones parameters. 

\begin{table}[h!]
\centering
\caption{ASFE values obtained with MM approach and three book-ending correction protocols (all values are in kcal/mol).}
\label{table1}
\begin{tabular}{lcccc c}
\hline\hline
\textbf{System} & 
\makecell{\textbf{MM} \\ \textbf{HFE}} & 
\makecell{\textbf{HF} \\ \textbf{protocol}} & 
\makecell{\textbf{HF + FCI} \\ \textbf{protocol}} & 
\makecell{\textbf{HF + SQD} \\ \textbf{protocol}} & 
\makecell{\textbf{MNSol}} \\
\hline
Ammonia & -3.87& -1.94& -2.35&   -2.20& -4.29\\
Methane &       2.28&       2.13&       1.44&     1.91& 2.00\\
Water   &       -8.96&       -7.02&       -7.58&      -7.44& -6.31\\
\hline\hline
\end{tabular}
\end{table}

For the first time, we demonstrate that quantum hardware can be integrated into classical alchemical workflows to enhance the accuracy of free energy predictions, alongside with the incorporation of an FCI solver. This marks a key step toward hybrid quantum-classical simulations. Despite the discrepancies observed, particularly for ammonia, the results reported in Table~\ref{table1} show reproducible and consistent corrections across three different QM approaches. The numerical differences observed between HF, HF + FCI, and HF + SQD approaches reflect their inherent methodological differences. Taken together, our results validate the hybrid QM/MM interface developed here, which enhances the HFE calculations by integrating different quantum backends within the same framework. Despite the observation that quantum corrections do not always yield improved values over classical MM results, likely because of the current selection of QM methods and/or Lennard-Jones parameters, the proposed interface is modular, extensible and supports dynamic switching between classical and more sophisticated quantum engines. Notably, this work represents the first embedding of a quantum hardware within alchemical workflows.

\section{Conclusions and Outlook}

AFE methods are widely used to estimate free energy differences associated with molecular transformations, which is essential for predicting properties like binding affinity. While quantum algorithms have been applied to general electronic structure calculations, their integration into AFE transformations remains limited. Here, we introduce quantum computing into AFE calculations, targeting this largely unexplored yet critical area, particularly relevant in fields such as drug discovery. \cite{devivo2016} 
Quantum algorithms specifically designed for free energy estimation represents a promising new direction for applying quantum hardware to chemistry related challenges. A key contribution of this work is the integration of configurations interaction methods, (FCI and SQD) with established classical tools, where previously book-ending correction was performed only with HF and DFT methods. Notably, the present work represents the first SQD study where the nuclear gradients were calculated with SQD method, while all previous SQD studies were addressing only the energies of the studied systems \cite{robledo2024chemistry, kaliakin2024accurate}. The proposed workflow connects the MD package AMBER to quantum backends supporting CI simulations, enabling seamless hybrid quantum-classical simulations of free energy calculations. This integration not only enables both conventional simulations on classic hardware and quantum-centric SQD simulations on real quantum device, but also highlights the synergy between the classical and quantum platforms. By tailoring quantum corrections specifically for AFE workflows, our approach opens the door to more accurate predictions and systematic validation and benchmarking of current databases \cite{Manerich2020}. In our benchmark tests on small molecules, the inclusion of CI-based book-ending corrections using quantum-centric methods yielded to free energy correction that moved our classical results closer to the MNSol database references, particularly for methane and water. However, the deviation of the corrected HFE values of ammonia highlight that important challenges remain. Despite these limitations, these results are encouraging and support the application of quantum-centric AFE workflows, as quantum hardware and algorithms continue to evolve. This enhanced level of accuracy will be particularly valuable for applications in drug discovery and lead optimization. Furthermore, the free energy perturbation schemes are naturally suited for parallel computation across multiple quantum devices, since the lambda windows of the production step are independent by design. Exploiting this inherent parallelism can significantly increase the utility of quantum devices, paving the way for scalable workflows in preparation for the post-NISQ era.

\begin{acknowledgement}

The authors gratefully acknowledge financial support from the National Science Foundation (NSF) through CSSI Frameworks Grant OAC-2209717 and from the National Institutes of Health (Grant Numbers GM130641). The authors also thank Javier Robledo Moreno, Mario Motta, Thaddeus Pellegrini, Caleb Johnson, Abdullah Ash Saki, Iskandar Sitdikov, Kevin Sung, and Antonio Mezzacapo for their guidance on the LUCJ ansatz and the SQD method.

\end{acknowledgement}

\section*{Code Availability}
Qiskit, ffsim, and Qiskit IBM Runtime utilized for LUCJ simulations can be obtained from corresponding GitHub repositories:

\url{https://github.com/qiskit-community/ffsim}

\url{https://github.com/Qiskit/qiskit}

\url{https://github.com/Qiskit/qiskit-ibm-runtime}

Configuration recovery code is distributed as Qiskit SQD addon:

\url{https://github.com/Qiskit/qiskit-addon-sqd}

PySCF through corresponding GitHub repository:

\url{https://github.com/pyscf/pyscf}

The tutorial demonstrating full SQD workflow is available below:

\url{https://qiskit.github.io/qiskit-addon-sqd/tutorials/01_chemistry_hamiltonian.html}

\section*{Competing Interest}
The authors declare no competing interest.


\bibliography{AFE_QPU}

\providecommand{\latin}[1]{#1}
\makeatletter
\providecommand{\doi}
  {\begingroup\let\do\@makeother\dospecials
  \catcode`\{=1 \catcode`\}=2 \doi@aux}
\providecommand{\doi@aux}[1]{\endgroup\texttt{#1}}
\makeatother
\providecommand*\mcitethebibliography{\thebibliography}
\csname @ifundefined\endcsname{endmcitethebibliography}  {\let\endmcitethebibliography\endthebibliography}{}
\begin{mcitethebibliography}{70}
\providecommand*\natexlab[1]{#1}
\providecommand*\mciteSetBstSublistMode[1]{}
\providecommand*\mciteSetBstMaxWidthForm[2]{}
\providecommand*\mciteBstWouldAddEndPuncttrue
  {\def\EndOfBibitem{\unskip.}}
\providecommand*\mciteBstWouldAddEndPunctfalse
  {\let\EndOfBibitem\relax}
\providecommand*\mciteSetBstMidEndSepPunct[3]{}
\providecommand*\mciteSetBstSublistLabelBeginEnd[3]{}
\providecommand*\EndOfBibitem{}
\mciteSetBstSublistMode{f}
\mciteSetBstMaxWidthForm{subitem}{(\alph{mcitesubitemcount})}
\mciteSetBstSublistLabelBeginEnd
  {\mcitemaxwidthsubitemform\space}
  {\relax}
  {\relax}

\bibitem[Straatsma and McCammon(1992)Straatsma, and McCammon]{afe}
Straatsma,~T.~P.; McCammon,~J.~A. Computational Alchemy. \emph{Annual Review of Physical Chemistry} \textbf{1992}, \emph{43}, 407--435\relax
\mciteBstWouldAddEndPuncttrue
\mciteSetBstMidEndSepPunct{\mcitedefaultmidpunct}
{\mcitedefaultendpunct}{\mcitedefaultseppunct}\relax
\EndOfBibitem
\bibitem[De~Vivo \latin{et~al.}(2016)De~Vivo, Masetti, Bottegoni, and Cavalli]{devivo2016}
De~Vivo,~M.; Masetti,~M.; Bottegoni,~G.; Cavalli,~A. Role of Molecular Dynamics and Related Methods in Drug Discovery. \emph{Journal of Medicinal Chemistry} \textbf{2016}, \emph{59}, 4035--4061, PMID: 26807648\relax
\mciteBstWouldAddEndPuncttrue
\mciteSetBstMidEndSepPunct{\mcitedefaultmidpunct}
{\mcitedefaultendpunct}{\mcitedefaultseppunct}\relax
\EndOfBibitem
\bibitem[Jorgensen and Duffy(2000)Jorgensen, and Duffy]{jorgensen2000}
Jorgensen,~W.~L.; Duffy,~E.~M. Prediction of drug solubility from Monte Carlo simulations. \emph{Bioorganic $\&$ Medicinal Chemistry Letters} \textbf{2000}, \emph{10}, 1155--1158\relax
\mciteBstWouldAddEndPuncttrue
\mciteSetBstMidEndSepPunct{\mcitedefaultmidpunct}
{\mcitedefaultendpunct}{\mcitedefaultseppunct}\relax
\EndOfBibitem
\bibitem[Li \latin{et~al.}(2024)Li, Yin, Li, Ma, Yi, Zhang, Zou, Bu, Dai, Yue, Chen, Zhang, and Zhang]{li2024}
Li,~W.; Yin,~Z.; Li,~X.; Ma,~D.; Yi,~S.; Zhang,~Z.; Zou,~C.; Bu,~K.; Dai,~M.; Yue,~J. \latin{et~al.}  A hybrid quantum computing pipeline for real world drug discovery. \emph{Scientific Reports} \textbf{2024}, \emph{14}\relax
\mciteBstWouldAddEndPuncttrue
\mciteSetBstMidEndSepPunct{\mcitedefaultmidpunct}
{\mcitedefaultendpunct}{\mcitedefaultseppunct}\relax
\EndOfBibitem
\bibitem[Salo-Ahen \latin{et~al.}(2021)Salo-Ahen, Alanko, Bhadane, Bonvin, Honorato, Hossain, Juffer, Kabedev, Lahtela-Kakkonen, Larsen, Lescrinier, Marimuthu, Mirza, Mustafa, Nunes-Alves, Pantsar, Saadabadi, Singaravelu, and Vanmeert]{saloahen2021}
Salo-Ahen,~O. M.~H.; Alanko,~I.; Bhadane,~R.; Bonvin,~A. M. J.~J.; Honorato,~R.~V.; Hossain,~S.; Juffer,~A.~H.; Kabedev,~A.; Lahtela-Kakkonen,~M.; Larsen,~A.~S. \latin{et~al.}  Molecular Dynamics Simulations in Drug Discovery and Pharmaceutical Development. \emph{Processes} \textbf{2021}, \emph{9}\relax
\mciteBstWouldAddEndPuncttrue
\mciteSetBstMidEndSepPunct{\mcitedefaultmidpunct}
{\mcitedefaultendpunct}{\mcitedefaultseppunct}\relax
\EndOfBibitem
\bibitem[Zeng \latin{et~al.}(2023)Zeng, Tao, Giese, and York]{zeng2023}
Zeng,~J.; Tao,~Y.; Giese,~T.~J.; York,~D.~M. QD$\pi$: A Quantum Deep Potential Interaction Model for Drug Discovery. \emph{Journal of Chemical Theory and Computation} \textbf{2023}, \emph{19}, 1261--1275, PMID: 36696673\relax
\mciteBstWouldAddEndPuncttrue
\mciteSetBstMidEndSepPunct{\mcitedefaultmidpunct}
{\mcitedefaultendpunct}{\mcitedefaultseppunct}\relax
\EndOfBibitem
\bibitem[Giese \latin{et~al.}(2022)Giese, Zeng, Ekesan, and York]{giese2022}
Giese,~T.~J.; Zeng,~J.; Ekesan,~{\c{S}\"olen}.; York,~D.~M. Combined QM/MM, Machine Learning Path Integral Approach to Compute Free Energy Profiles and Kinetic Isotope Effects in RNA Cleavage Reactions. \emph{Journal of Chemical Theory and Computation} \textbf{2022}, \emph{18}, 4304--4317, PMID: 35709391\relax
\mciteBstWouldAddEndPuncttrue
\mciteSetBstMidEndSepPunct{\mcitedefaultmidpunct}
{\mcitedefaultendpunct}{\mcitedefaultseppunct}\relax
\EndOfBibitem
\bibitem[Clemente \latin{et~al.}(2023)Clemente, Capece, and Martí]{clemente2023}
Clemente,~C.~M.; Capece,~L.; Martí,~M.~A. Best Practices on QM/MM Simulations of Biological Systems. \emph{Journal of Chemical Information and Modeling} \textbf{2023}, \emph{63}, 2609--2627, PMID: 37100031\relax
\mciteBstWouldAddEndPuncttrue
\mciteSetBstMidEndSepPunct{\mcitedefaultmidpunct}
{\mcitedefaultendpunct}{\mcitedefaultseppunct}\relax
\EndOfBibitem
\bibitem[Kar(2023)]{Kar2023}
Kar,~R.~K. Benefits of hybrid QM/MM over traditional classical mechanics in pharmaceutical systems. \emph{Drug Discovery Today} \textbf{2023}, \emph{28}, 103374\relax
\mciteBstWouldAddEndPuncttrue
\mciteSetBstMidEndSepPunct{\mcitedefaultmidpunct}
{\mcitedefaultendpunct}{\mcitedefaultseppunct}\relax
\EndOfBibitem
\bibitem[Warshel and Levitt(1976)Warshel, and Levitt]{warshel1976}
Warshel,~A.; Levitt,~M. Theoretical studies of enzymic reactions: Dielectric, electrostatic and steric stabilization of the carbonium ion in the reaction of lysozyme. \emph{Journal of Molecular Biology} \textbf{1976}, \emph{103}, 227--249\relax
\mciteBstWouldAddEndPuncttrue
\mciteSetBstMidEndSepPunct{\mcitedefaultmidpunct}
{\mcitedefaultendpunct}{\mcitedefaultseppunct}\relax
\EndOfBibitem
\bibitem[Duarte \latin{et~al.}(2015)Duarte, Amrein, Blaha-Nelson, and Kamerlin]{duarte2015}
Duarte,~F.; Amrein,~B.~A.; Blaha-Nelson,~D.; Kamerlin,~S.~C. Recent advances in QM/MM free energy calculations using reference potentials. \emph{Biochimica et Biophysica Acta (BBA) - General Subjects} \textbf{2015}, \emph{1850}, 954--965, Recent developments of molecular dynamics\relax
\mciteBstWouldAddEndPuncttrue
\mciteSetBstMidEndSepPunct{\mcitedefaultmidpunct}
{\mcitedefaultendpunct}{\mcitedefaultseppunct}\relax
\EndOfBibitem
\bibitem[Manathunga \latin{et~al.}(2022)Manathunga, Götz, and Merz]{manathunga2022}
Manathunga,~M.; Götz,~A.~W.; Merz,~K.~M. Computer-aided drug design, quantum-mechanical methods for biological problems. \emph{Current Opinion in Structural Biology} \textbf{2022}, \emph{75}, 102417\relax
\mciteBstWouldAddEndPuncttrue
\mciteSetBstMidEndSepPunct{\mcitedefaultmidpunct}
{\mcitedefaultendpunct}{\mcitedefaultseppunct}\relax
\EndOfBibitem
\bibitem[Lu \latin{et~al.}(2016)Lu, Fang, Ito, Okamoto, Ovchinnikov, and and]{Lu01092016}
Lu,~X.; Fang,~D.; Ito,~S.; Okamoto,~Y.; Ovchinnikov,~V.; and,~Q.~C. QM/MM free energy simulations: recent progress and challenges. \emph{Molecular Simulation} \textbf{2016}, \emph{42}, 1056--1078, PMID: 27563170\relax
\mciteBstWouldAddEndPuncttrue
\mciteSetBstMidEndSepPunct{\mcitedefaultmidpunct}
{\mcitedefaultendpunct}{\mcitedefaultseppunct}\relax
\EndOfBibitem
\bibitem[Giese \latin{et~al.}(2024)Giese, Zeng, Lerew, McCarthy, Tao, Ekesan, and York]{giese2024}
Giese,~T.~J.; Zeng,~J.; Lerew,~L.; McCarthy,~E.; Tao,~Y.; Ekesan,~{\c{S}\"olen}.; York,~D.~M. Software Infrastructure for Next-Generation QM/MM-$\triangle$MLP Force Fields. \emph{The Journal of Physical Chemistry B} \textbf{2024}, \emph{128}, 6257--6271, PMID: 38905451\relax
\mciteBstWouldAddEndPuncttrue
\mciteSetBstMidEndSepPunct{\mcitedefaultmidpunct}
{\mcitedefaultendpunct}{\mcitedefaultseppunct}\relax
\EndOfBibitem
\bibitem[York(2023)]{york2023}
York,~D.~M. Modern Alchemical Free Energy Methods for Drug Discovery Explained. \emph{ACS Physical Chemistry Au} \textbf{2023}, \emph{3}, 478--491\relax
\mciteBstWouldAddEndPuncttrue
\mciteSetBstMidEndSepPunct{\mcitedefaultmidpunct}
{\mcitedefaultendpunct}{\mcitedefaultseppunct}\relax
\EndOfBibitem
\bibitem[Li \latin{et~al.}(2003)Li, Zhang, and Cui]{Li2003}
Li,~G.; Zhang,~X.; Cui,~Q. Free energy perturbation calculations with combined QM/MM potentials: complications, simplifications, and applications to redox potential calculations. \emph{J. Phys. Chem. B} \textbf{2003}, \emph{107}, 8643--8653\relax
\mciteBstWouldAddEndPuncttrue
\mciteSetBstMidEndSepPunct{\mcitedefaultmidpunct}
{\mcitedefaultendpunct}{\mcitedefaultseppunct}\relax
\EndOfBibitem
\bibitem[Gao(1992)]{Gao1992}
Gao,~J. Absolute free energy of solvation from Monte Carlo simulations using combined quantum and molecular mechanical potentials. \emph{The Journal of Physical Chemistry} \textbf{1992}, \emph{96}, 537--540\relax
\mciteBstWouldAddEndPuncttrue
\mciteSetBstMidEndSepPunct{\mcitedefaultmidpunct}
{\mcitedefaultendpunct}{\mcitedefaultseppunct}\relax
\EndOfBibitem
\bibitem[Gao and Xia(1992)Gao, and Xia]{gao1992priori}
Gao,~J.; Xia,~X. A priori evaluation of aqueous polarization effects through Monte Carlo QM-MM simulations. \emph{Science} \textbf{1992}, \emph{258}, 631--635\relax
\mciteBstWouldAddEndPuncttrue
\mciteSetBstMidEndSepPunct{\mcitedefaultmidpunct}
{\mcitedefaultendpunct}{\mcitedefaultseppunct}\relax
\EndOfBibitem
\bibitem[{Luzhkov} and {Warshel}(1992){Luzhkov}, and {Warshel}]{Warshel1992}
{Luzhkov},~V.; {Warshel},~A. {Microscopic models for quantum mechanical calculations of chemical processes in solutions: LD/AMPAC and SCAAS/AMPAC calculations of solvation energies}. \emph{Journal of Computational Chemistry} \textbf{1992}, \emph{13}, 199--213\relax
\mciteBstWouldAddEndPuncttrue
\mciteSetBstMidEndSepPunct{\mcitedefaultmidpunct}
{\mcitedefaultendpunct}{\mcitedefaultseppunct}\relax
\EndOfBibitem
\bibitem[Kearns \latin{et~al.}(2019)Kearns, Warrensford, Boresch, and Woodcock]{molecules24040681}
Kearns,~F.~L.; Warrensford,~L.; Boresch,~S.; Woodcock,~H.~L. The Good, the Bad, and the Ugly: “HiPen”, a New Dataset for Validating (S)QM/MM Free Energy Simulations. \emph{Molecules} \textbf{2019}, \emph{24}\relax
\mciteBstWouldAddEndPuncttrue
\mciteSetBstMidEndSepPunct{\mcitedefaultmidpunct}
{\mcitedefaultendpunct}{\mcitedefaultseppunct}\relax
\EndOfBibitem
\bibitem[Hudson \latin{et~al.}(2018)Hudson, Han, Woodcock, and Brooks]{hudson2018force}
Hudson,~P.~S.; Han,~K.; Woodcock,~H.~L.; Brooks,~B.~R. Force matching as a stepping stone to QM/MM CB[8] host/guest binding free energies: a SAMPL6 cautionary tale. \emph{Journal of Computer-Aided Molecular Design} \textbf{2018}, \emph{32}, 983--999\relax
\mciteBstWouldAddEndPuncttrue
\mciteSetBstMidEndSepPunct{\mcitedefaultmidpunct}
{\mcitedefaultendpunct}{\mcitedefaultseppunct}\relax
\EndOfBibitem
\bibitem[Hudson \latin{et~al.}(2018)Hudson, Boresch, Rogers, and Woodcock]{hudson2018accelerating}
Hudson,~P.~S.; Boresch,~S.; Rogers,~D.~M.; Woodcock,~H.~L. Accelerating QM/MM Free Energy Computations via Intramolecular Force Matching. \emph{Journal of Chemical Theory and Computation} \textbf{2018}, \emph{14}, 6327--6335\relax
\mciteBstWouldAddEndPuncttrue
\mciteSetBstMidEndSepPunct{\mcitedefaultmidpunct}
{\mcitedefaultendpunct}{\mcitedefaultseppunct}\relax
\EndOfBibitem
\bibitem[Kearns \latin{et~al.}(2017)Kearns, Hudson, Woodcock, and Boresch]{Kearns2017}
Kearns,~F.~L.; Hudson,~P.~S.; Woodcock,~H.~L.; Boresch,~S. Computing converged free energy differences between levels of theory via nonequilibrium work methods: Challenges and opportunities. \emph{Journal of Computational Chemistry} \textbf{2017}, \emph{38}, 1376--1388\relax
\mciteBstWouldAddEndPuncttrue
\mciteSetBstMidEndSepPunct{\mcitedefaultmidpunct}
{\mcitedefaultendpunct}{\mcitedefaultseppunct}\relax
\EndOfBibitem
\bibitem[König \latin{et~al.}(2014)König, Hudson, Boresch, and Woodcock]{konig2014}
König,~G.; Hudson,~P.~S.; Boresch,~S.; Woodcock,~H.~L. Multiscale Free Energy Simulations: An Efficient Method for Connecting Classical MD Simulations to QM or QM/MM Free Energies Using Non-Boltzmann Bennett Reweighting Schemes. \emph{Journal of Chemical Theory and Computation} \textbf{2014}, \emph{10}, 1406--1419, PMID: 24803863\relax
\mciteBstWouldAddEndPuncttrue
\mciteSetBstMidEndSepPunct{\mcitedefaultmidpunct}
{\mcitedefaultendpunct}{\mcitedefaultseppunct}\relax
\EndOfBibitem
\bibitem[König \latin{et~al.}(2014)König, Hudson, Boresch, and Woodcock]{konig2014multiscale}
König,~G.; Hudson,~P.~S.; Boresch,~S.; Woodcock,~H.~L. Multiscale Free Energy Simulations: An Efficient Method for Connecting Classical MD Simulations to QM or QM/MM Free Energies Using Non-Boltzmann Bennett Reweighting Schemes. \emph{Journal of Chemical Theory and Computation} \textbf{2014}, \emph{10}, 1406--1419, PMID: 24803863\relax
\mciteBstWouldAddEndPuncttrue
\mciteSetBstMidEndSepPunct{\mcitedefaultmidpunct}
{\mcitedefaultendpunct}{\mcitedefaultseppunct}\relax
\EndOfBibitem
\bibitem[König and Brooks(2015)König, and Brooks]{konig2015}
König,~G.; Brooks,~B.~R. Correcting for the free energy costs of bond or angle constraints in molecular dynamics simulations. \emph{Biochimica et Biophysica Acta (BBA) - General Subjects} \textbf{2015}, \emph{1850}, 932--943\relax
\mciteBstWouldAddEndPuncttrue
\mciteSetBstMidEndSepPunct{\mcitedefaultmidpunct}
{\mcitedefaultendpunct}{\mcitedefaultseppunct}\relax
\EndOfBibitem
\bibitem[König \latin{et~al.}(2018)König, Brooks, Thiel, and York]{konig2018}
König,~G.; Brooks,~B.~R.; Thiel,~W.; York,~D.~M. On the convergence of multi-scale free energy simulations. \emph{Molecular Simulation} \textbf{2018}, \emph{44}, 1062--1081\relax
\mciteBstWouldAddEndPuncttrue
\mciteSetBstMidEndSepPunct{\mcitedefaultmidpunct}
{\mcitedefaultendpunct}{\mcitedefaultseppunct}\relax
\EndOfBibitem
\bibitem[König \latin{et~al.}(2018)König, Pickard, Huang, Thiel, MacKerell, Brooks, and York]{konig2018comparison}
König,~G.; Pickard,~F.~C.; Huang,~J.; Thiel,~W.; MacKerell,~A.~D.; Brooks,~B.~R.; York,~D.~M. A Comparison of QM/MM Simulations with and without the Drude Oscillator Model Based on Hydration Free Energies of Simple Solutes. \emph{Molecules} \textbf{2018}, \emph{23}, 2695\relax
\mciteBstWouldAddEndPuncttrue
\mciteSetBstMidEndSepPunct{\mcitedefaultmidpunct}
{\mcitedefaultendpunct}{\mcitedefaultseppunct}\relax
\EndOfBibitem
\bibitem[Hudson \latin{et~al.}(2015)Hudson, Woodcock, and Boresch]{hudson2015}
Hudson,~P.~S.; Woodcock,~H.~L.; Boresch,~S. Use of Nonequilibrium Work Methods to Compute Free Energy Differences Between Molecular Mechanical and Quantum Mechanical Representations of Molecular Systems. \emph{The Journal of Physical Chemistry Letters} \textbf{2015}, \emph{6}, 4850--4856, PMID: 26539729\relax
\mciteBstWouldAddEndPuncttrue
\mciteSetBstMidEndSepPunct{\mcitedefaultmidpunct}
{\mcitedefaultendpunct}{\mcitedefaultseppunct}\relax
\EndOfBibitem
\bibitem[Giese and York(2019)Giese, and York]{giese2019}
Giese,~T.~J.; York,~D.~M. Development of a Robust Indirect Approach for MM → QM Free Energy Calculations That Combines Force-Matched Reference Potential and Bennett’s Acceptance Ratio Methods. \emph{Journal of Chemical Theory and Computation} \textbf{2019}, \emph{15}, 5543--5562, PMID: 31507179\relax
\mciteBstWouldAddEndPuncttrue
\mciteSetBstMidEndSepPunct{\mcitedefaultmidpunct}
{\mcitedefaultendpunct}{\mcitedefaultseppunct}\relax
\EndOfBibitem
\bibitem[Cohen \latin{et~al.}(2012)Cohen, Mori-Sánchez, and Yang]{dftrew}
Cohen,~A.~J.; Mori-Sánchez,~P.; Yang,~W. Challenges for Density Functional Theory. \emph{Chemical Reviews} \textbf{2012}, \emph{112}, 289--320, PMID: 22191548\relax
\mciteBstWouldAddEndPuncttrue
\mciteSetBstMidEndSepPunct{\mcitedefaultmidpunct}
{\mcitedefaultendpunct}{\mcitedefaultseppunct}\relax
\EndOfBibitem
\bibitem[Gao \latin{et~al.}(2024)Gao, Imamura, Kasagi, and Yoshida]{Gao2024}
Gao,~H.; Imamura,~S.; Kasagi,~A.; Yoshida,~E. Distributed Implementation of Full Configuration Interaction for One Trillion Determinants. \emph{Journal of Chemical Theory and Computation} \textbf{2024}, \emph{20}, 1185--1192, PMID: 38314701\relax
\mciteBstWouldAddEndPuncttrue
\mciteSetBstMidEndSepPunct{\mcitedefaultmidpunct}
{\mcitedefaultendpunct}{\mcitedefaultseppunct}\relax
\EndOfBibitem
\bibitem[Vogiatzis \latin{et~al.}(2017)Vogiatzis, Ma, Olsen, Gagliardi, and de~Jong]{Vogiatzis2017}
Vogiatzis,~K.~D.; Ma,~D.; Olsen,~J.; Gagliardi,~L.; de~Jong,~W.~A. {Pushing configuration-interaction to the limit: Towards massively parallel MCSCF calculations}. \emph{The Journal of Chemical Physics} \textbf{2017}, \emph{147}, 184111\relax
\mciteBstWouldAddEndPuncttrue
\mciteSetBstMidEndSepPunct{\mcitedefaultmidpunct}
{\mcitedefaultendpunct}{\mcitedefaultseppunct}\relax
\EndOfBibitem
\bibitem[Abraham and Mayhall(2020)Abraham, and Mayhall]{Abraham}
Abraham,~V.; Mayhall,~N.~J. Selected Configuration Interaction in a Basis of Cluster State Tensor Products. \emph{Journal of Chemical Theory and Computation} \textbf{2020}, \emph{16}, 6098--6113, PMID: 32846094\relax
\mciteBstWouldAddEndPuncttrue
\mciteSetBstMidEndSepPunct{\mcitedefaultmidpunct}
{\mcitedefaultendpunct}{\mcitedefaultseppunct}\relax
\EndOfBibitem
\bibitem[Sharma \latin{et~al.}(2017)Sharma, Holmes, Jeanmairet, Alavi, and Umrigar]{sharma2017semistochastic}
Sharma,~S.; Holmes,~A.~A.; Jeanmairet,~G.; Alavi,~A.; Umrigar,~C.~J. Semistochastic heat-bath configuration interaction method: Selected configuration interaction with semistochastic perturbation theory. \emph{Journal of Chemical Theory and Computation} \textbf{2017}, \emph{13}, 1595--1604\relax
\mciteBstWouldAddEndPuncttrue
\mciteSetBstMidEndSepPunct{\mcitedefaultmidpunct}
{\mcitedefaultendpunct}{\mcitedefaultseppunct}\relax
\EndOfBibitem
\bibitem[Holmes \latin{et~al.}(2016)Holmes, Tubman, and Umrigar]{holmes2016heat}
Holmes,~A.~A.; Tubman,~N.~M.; Umrigar,~C. Heat-bath configuration interaction: An efficient selected configuration interaction algorithm inspired by heat-bath sampling. \emph{Journal of Chemical Theory and Computation} \textbf{2016}, \emph{12}, 3674--3680\relax
\mciteBstWouldAddEndPuncttrue
\mciteSetBstMidEndSepPunct{\mcitedefaultmidpunct}
{\mcitedefaultendpunct}{\mcitedefaultseppunct}\relax
\EndOfBibitem
\bibitem[Robledo-Moreno \latin{et~al.}(2025)Robledo-Moreno, Motta, Haas, Javadi-Abhari, Jurcevic, Kirby, Martiel, Sharma, Sharma, Shirakawa, Sitdikov, Sun, Sung, Takita, Tran, Yunoki, and Mezzacapo]{robledo2024chemistry}
Robledo-Moreno,~J.; Motta,~M.; Haas,~H.; Javadi-Abhari,~A.; Jurcevic,~P.; Kirby,~W.; Martiel,~S.; Sharma,~K.; Sharma,~S.; Shirakawa,~T. \latin{et~al.}  Chemistry beyond the scale of exact diagonalization on a quantum-centric supercomputer. \emph{Science Advances} \textbf{2025}, \emph{11}, eadu9991\relax
\mciteBstWouldAddEndPuncttrue
\mciteSetBstMidEndSepPunct{\mcitedefaultmidpunct}
{\mcitedefaultendpunct}{\mcitedefaultseppunct}\relax
\EndOfBibitem
\bibitem[Shajan \latin{et~al.}(2024)Shajan, Kaliakin, Mitra, Moreno, Li, Motta, Johnson, Saki, Das, Sitdikov, \latin{et~al.} others]{shajan2024towards}
Shajan,~A.; Kaliakin,~D.; Mitra,~A.; Moreno,~J.~R.; Li,~Z.; Motta,~M.; Johnson,~C.; Saki,~A.~A.; Das,~S.; Sitdikov,~I. \latin{et~al.}  Towards quantum-centric simulations of extended molecules: sample-based quantum diagonalization enhanced with density matrix embedding theory. \emph{arXiv preprint arXiv:2411.09861} \textbf{2024}, \relax
\mciteBstWouldAddEndPunctfalse
\mciteSetBstMidEndSepPunct{\mcitedefaultmidpunct}
{}{\mcitedefaultseppunct}\relax
\EndOfBibitem
\bibitem[Liepuoniute \latin{et~al.}(2024)Liepuoniute, Doney, Robledo-Moreno, Job, Friend, and Jones]{Liepuoniute2024}
Liepuoniute,~I.; Doney,~K.~D.; Robledo-Moreno,~J.; Job,~J.~A.; Friend,~W.~S.; Jones,~G.~O. Quantum-Centric Study of Methylene Singlet and Triplet States. \textbf{2024}, \relax
\mciteBstWouldAddEndPunctfalse
\mciteSetBstMidEndSepPunct{\mcitedefaultmidpunct}
{}{\mcitedefaultseppunct}\relax
\EndOfBibitem
\bibitem[Barison \latin{et~al.}(2025)Barison, Robledo~Moreno, and Motta]{Barison2025}
Barison,~S.; Robledo~Moreno,~J.; Motta,~M. Quantum-centric computation of molecular excited states with extended sample-based quantum diagonalization. \emph{Quantum Science and Technology} \textbf{2025}, \emph{10}, 025034\relax
\mciteBstWouldAddEndPuncttrue
\mciteSetBstMidEndSepPunct{\mcitedefaultmidpunct}
{\mcitedefaultendpunct}{\mcitedefaultseppunct}\relax
\EndOfBibitem
\bibitem[Yu \latin{et~al.}(2025)Yu, Moreno, Iosue, Bertels, Claudino, Fuller, Groszkowski, Humble, Jurcevic, Kirby, Maier, Motta, Pokharel, Seif, Shehata, Sung, Tran, Tripathi, Mezzacapo, and Sharma]{yu2025quantum}
Yu,~J.; Moreno,~J.~R.; Iosue,~J.~T.; Bertels,~L.; Claudino,~D.; Fuller,~B.; Groszkowski,~P.; Humble,~T.~S.; Jurcevic,~P.; Kirby,~W. \latin{et~al.}  Quantum-Centric Algorithm for Sample-Based Krylov Diagonalization. \emph{arXiv preprint arXiv:2501.09702} \textbf{2025}, Submitted on 16 Jan 2025 (v1), last revised 24 Jan 2025 (this version, v2)\relax
\mciteBstWouldAddEndPuncttrue
\mciteSetBstMidEndSepPunct{\mcitedefaultmidpunct}
{\mcitedefaultendpunct}{\mcitedefaultseppunct}\relax
\EndOfBibitem
\bibitem[Kaliakin \latin{et~al.}(2025)Kaliakin, Shajan, Liang, and Merz]{kaliakin2025implicit}
Kaliakin,~D.; Shajan,~A.; Liang,~F.; Merz,~K. M.~J. Implicit Solvent Sample-Based Quantum Diagonalization. \emph{The Journal of Physical Chemistry B} \textbf{2025}, \emph{129}, 5788--5796\relax
\mciteBstWouldAddEndPuncttrue
\mciteSetBstMidEndSepPunct{\mcitedefaultmidpunct}
{\mcitedefaultendpunct}{\mcitedefaultseppunct}\relax
\EndOfBibitem
\bibitem[Raghavachari and Anderson(1996)Raghavachari, and Anderson]{Raghavachari}
Raghavachari,~K.; Anderson,~J.~B. Electron Correlation Effects in Molecules. \emph{The Journal of Physical Chemistry} \textbf{1996}, \emph{100}, 12960--12973\relax
\mciteBstWouldAddEndPuncttrue
\mciteSetBstMidEndSepPunct{\mcitedefaultmidpunct}
{\mcitedefaultendpunct}{\mcitedefaultseppunct}\relax
\EndOfBibitem
\bibitem[Martin(2022)]{Martin}
Martin,~J. M.~L. Electron Correlation: Nature's Weird and Wonderful Chemical Glue. \emph{Israel Journal of Chemistry} \textbf{2022}, \emph{62}, e202100111\relax
\mciteBstWouldAddEndPuncttrue
\mciteSetBstMidEndSepPunct{\mcitedefaultmidpunct}
{\mcitedefaultendpunct}{\mcitedefaultseppunct}\relax
\EndOfBibitem
\bibitem[Case \latin{et~al.}(2024)Case, Aktulga, Belfon, Cerutti, Cisneros, Cruzeiro, Forouzesh, Giese, G{\"o}tz, Gohlke, Izadi, Kasavajhala, Kaymak, King, Kurtzman, Lee, Li, Li, Liu, Luchko, Luo, Manathunga, Machado, Nguyen, O’Hearn, Onufriev, Pan, Pantano, Qi, Rahnamoun, Risheh, Schott-Verdugo, Shajan, Swails, Wang, Wei, Wu, Wu, Zhang, Zhao, Zhu, Cheatham, Roe, Roitberg, Simmerling, York, Nagan, and Merz]{amber24}
Case,~D.~A.; Aktulga,~H.~M.; Belfon,~K.; Cerutti,~D.~S.; Cisneros,~G.~A.; Cruzeiro,~V. W.~D.; Forouzesh,~N.; Giese,~T.~J.; G{\"o}tz,~A.~W.; Gohlke,~H. \latin{et~al.}  AMBER24. 2024; \url{https://ambermd.org}\relax
\mciteBstWouldAddEndPuncttrue
\mciteSetBstMidEndSepPunct{\mcitedefaultmidpunct}
{\mcitedefaultendpunct}{\mcitedefaultseppunct}\relax
\EndOfBibitem
\bibitem[He \latin{et~al.}(2020)He, Man, Yang, Lee, and Wang]{wang2020gaff2}
He,~X.; Man,~V.~H.; Yang,~W.; Lee,~T.-S.; Wang,~J. A fast and high-quality charge model for the next generation general AMBER force field. \emph{The Journal of Chemical Physics} \textbf{2020}, \emph{153}, 114502\relax
\mciteBstWouldAddEndPuncttrue
\mciteSetBstMidEndSepPunct{\mcitedefaultmidpunct}
{\mcitedefaultendpunct}{\mcitedefaultseppunct}\relax
\EndOfBibitem
\bibitem[Woods and Chappelle(2000)Woods, and Chappelle]{woods2000resp}
Woods,~R.; Chappelle,~R. Restrained electrostatic potential atomic partial charges for condensed-phase simulations of carbohydrates. \emph{Journal of Molecular Structure: THEOCHEM} \textbf{2000}, \emph{527}, 149--156\relax
\mciteBstWouldAddEndPuncttrue
\mciteSetBstMidEndSepPunct{\mcitedefaultmidpunct}
{\mcitedefaultendpunct}{\mcitedefaultseppunct}\relax
\EndOfBibitem
\bibitem[Frisch \latin{et~al.}(2016)Frisch, Trucks, Schlegel, Scuseria, Robb, Cheeseman, Scalmani, Barone, Petersson, Nakatsuji, Li, Caricato, Marenich, Bloino, Janesko, Gomperts, Mennucci, Hratchian, Ortiz, Izmaylov, Sonnenberg, Williams-Young, Ding, Lipparini, Egidi, Goings, Peng, Petrone, Henderson, Ranasinghe, Zakrzewski, Gao, Rega, Zheng, Liang, Hada, Ehara, Toyota, Fukuda, Hasegawa, Ishida, Nakajima, Honda, Kitao, Nakai, Vreven, Throssell, Montgomery, Peralta, Ogliaro, Bearpark, Heyd, Brothers, Kudin, Staroverov, Keith, Kobayashi, Normand, Raghavachari, Rendell, Burant, Iyengar, Tomasi, Cossi, Millam, Klene, Adamo, Cammi, Ochterski, Martin, Morokuma, Farkas, Foresman, and Fox]{g16}
Frisch,~M.~J.; Trucks,~G.~W.; Schlegel,~H.~B.; Scuseria,~G.~E.; Robb,~M.~A.; Cheeseman,~J.~R.; Scalmani,~G.; Barone,~V.; Petersson,~G.~A.; Nakatsuji,~H. \latin{et~al.}  Gaussian˜16 {R}evision {C}.01. \textbf{2016}, Gaussian Inc. Wallingford CT\relax
\mciteBstWouldAddEndPuncttrue
\mciteSetBstMidEndSepPunct{\mcitedefaultmidpunct}
{\mcitedefaultendpunct}{\mcitedefaultseppunct}\relax
\EndOfBibitem
\bibitem[Izadi and Onufriev(2016)Izadi, and Onufriev]{onufriev2016opc3}
Izadi,~S.; Onufriev,~A.~V. Accuracy limit of rigid 3-point water models. \emph{The Journal of Chemical Physics} \textbf{2016}, \emph{145}, 074501, PMID: 27544113\relax
\mciteBstWouldAddEndPuncttrue
\mciteSetBstMidEndSepPunct{\mcitedefaultmidpunct}
{\mcitedefaultendpunct}{\mcitedefaultseppunct}\relax
\EndOfBibitem
\bibitem[Izadi \latin{et~al.}(2014)Izadi, Anandakrishnan, and Onufriev]{onufriev2014opc}
Izadi,~S.; Anandakrishnan,~R.; Onufriev,~A.~V. Building Water Models: A Different Approach. \emph{The Journal of Physical Chemistry Letters} \textbf{2014}, \emph{5}, 3863--3871, PMID: 25400877\relax
\mciteBstWouldAddEndPuncttrue
\mciteSetBstMidEndSepPunct{\mcitedefaultmidpunct}
{\mcitedefaultendpunct}{\mcitedefaultseppunct}\relax
\EndOfBibitem
\bibitem[Darden \latin{et~al.}(1993)Darden, York, and Pedersen]{darden1993pme}
Darden,~T.; York,~D.; Pedersen,~L. Particle mesh Ewald: An N$\cdot$log(N) method for Ewald sums in large systems. \emph{The Journal of Chemical Physics} \textbf{1993}, \emph{98}, 10089--10092\relax
\mciteBstWouldAddEndPuncttrue
\mciteSetBstMidEndSepPunct{\mcitedefaultmidpunct}
{\mcitedefaultendpunct}{\mcitedefaultseppunct}\relax
\EndOfBibitem
\bibitem[Berendsen \latin{et~al.}(1984)Berendsen, Postma, van Gunsteren, DiNola, and Haak]{berendsen1984bath}
Berendsen,~H. J.~C.; Postma,~J. P.~M.; van Gunsteren,~W.~F.; DiNola,~A.; Haak,~J.~R. Molecular dynamics with coupling to an external bath. \emph{The Journal of Chemical Physics} \textbf{1984}, \emph{81}, 3684--3690\relax
\mciteBstWouldAddEndPuncttrue
\mciteSetBstMidEndSepPunct{\mcitedefaultmidpunct}
{\mcitedefaultendpunct}{\mcitedefaultseppunct}\relax
\EndOfBibitem
\bibitem[Hopkins \latin{et~al.}(2015)Hopkins, Le~Grand, Walker, and Roitberg]{hmr}
Hopkins,~C.~W.; Le~Grand,~S.; Walker,~R.~C.; Roitberg,~A.~E. Long-Time-Step Molecular Dynamics through Hydrogen Mass Repartitioning. \emph{Journal of Chemical Theory and Computation} \textbf{2015}, \emph{11}, 1864--1874\relax
\mciteBstWouldAddEndPuncttrue
\mciteSetBstMidEndSepPunct{\mcitedefaultmidpunct}
{\mcitedefaultendpunct}{\mcitedefaultseppunct}\relax
\EndOfBibitem
\bibitem[Manathunga \latin{et~al.}(2023)Manathunga, Aktulga, Götz, and Merz]{manathunga2023quantum}
Manathunga,~M.; Aktulga,~H.~M.; Götz,~A.~W.; Merz,~K.~M. Quantum Mechanics/Molecular Mechanics Simulations on NVIDIA and AMD Graphics Processing Units. \emph{J. Chem. Inf. Model.} \textbf{2023}, \emph{63}, 711--717\relax
\mciteBstWouldAddEndPuncttrue
\mciteSetBstMidEndSepPunct{\mcitedefaultmidpunct}
{\mcitedefaultendpunct}{\mcitedefaultseppunct}\relax
\EndOfBibitem
\bibitem[Cruzeiro \latin{et~al.}(2021)Cruzeiro, Manathunga, Merz, and Götz]{cruzeiro2021open}
Cruzeiro,~V. W.~D.; Manathunga,~M.; Merz,~K.~M.; Götz,~A.~W. Open-Source Multi-GPU-Accelerated QM/MM Simulations with AMBER and QUICK. \emph{J. Chem. Inf. Model.} \textbf{2021}, \emph{61}, 2109--2115\relax
\mciteBstWouldAddEndPuncttrue
\mciteSetBstMidEndSepPunct{\mcitedefaultmidpunct}
{\mcitedefaultendpunct}{\mcitedefaultseppunct}\relax
\EndOfBibitem
\bibitem[Shirts and Chodera(2008)Shirts, and Chodera]{mbar}
Shirts,~M.~R.; Chodera,~J.~D. Statistically optimal analysis of samples from multiple equilibrium states. \emph{The Journal of Chemical Physics} \textbf{2008}, \emph{129}, 124105\relax
\mciteBstWouldAddEndPuncttrue
\mciteSetBstMidEndSepPunct{\mcitedefaultmidpunct}
{\mcitedefaultendpunct}{\mcitedefaultseppunct}\relax
\EndOfBibitem
\bibitem[Aleksandrowicz \latin{et~al.}(2019)Aleksandrowicz, Alexander, Barkoutsos, Bello, Ben-Haim, Bucher, Cabrera-Hernández, Carballo-Franquis, Chen, Chen, Chow, Córcoles-Gonzales, Cross, Cross, Cruz-Benito, Culver, González, Torre, Ding, Dumitrescu, Duran, Eendebak, Everitt, Sertage, Frisch, Fuhrer, Gambetta, Gago, Gomez-Mosquera, Greenberg, Hamamura, Havlicek, Hellmers, Łukasz Herok, Horii, Hu, Imamichi, Itoko, Javadi-Abhari, Kanazawa, Karazeev, Krsulich, Liu, Luh, Maeng, Marques, Martín-Fernández, McClure, McKay, Meesala, Mezzacapo, Moll, Rodríguez, Nannicini, Nation, Ollitrault, O'Riordan, Paik, Pérez, Phan, Pistoia, Prutyanov, Reuter, Rice, Davila, Rudy, Ryu, Sathaye, Schnabel, Schoute, Setia, Shi, Silva, Siraichi, Sivarajah, Smolin, Soeken, Takahashi, Tavernelli, Taylor, Taylour, Trabing, Treinish, Turner, Vogt-Lee, Vuillot, Wildstrom, Wilson, Winston, Wood, Wood, Wörner, Akhalwaya, and Zoufal]{aleksandrowicz2019qiskit}
Aleksandrowicz,~G.; Alexander,~T.; Barkoutsos,~P.~K.; Bello,~L.; Ben-Haim,~Y.; Bucher,~D.; Cabrera-Hernández,~F.~J.; Carballo-Franquis,~J.; Chen,~A.; Chen,~C.-F. \latin{et~al.}  Qiskit: An Open-source Framework for Quantum Computing. 2019\relax
\mciteBstWouldAddEndPuncttrue
\mciteSetBstMidEndSepPunct{\mcitedefaultmidpunct}
{\mcitedefaultendpunct}{\mcitedefaultseppunct}\relax
\EndOfBibitem
\bibitem[Javadi-Abhari \latin{et~al.}(2024)Javadi-Abhari, Treinish, Krsulich, Wood, Lishman, Gacon, Martiel, Nation, Bishop, and Cross]{javadi2024quantum}
Javadi-Abhari,~A.; Treinish,~M.; Krsulich,~K.; Wood,~C.~J.; Lishman,~J.; Gacon,~J.; Martiel,~S.; Nation,~P.~D.; Bishop,~L.~S.; Cross,~A.~W. Quantum computing with Qiskit. \emph{arXiv preprint arXiv:2405.08810} \textbf{2024}, Available at \url{https://arxiv.org/abs/2405.08810}\relax
\mciteBstWouldAddEndPuncttrue
\mciteSetBstMidEndSepPunct{\mcitedefaultmidpunct}
{\mcitedefaultendpunct}{\mcitedefaultseppunct}\relax
\EndOfBibitem
\bibitem[Kaliakin \latin{et~al.}(2024)Kaliakin, Shajan, Moreno, Li, Mitra, Motta, Johnson, Saki, Das, and Sitdikov]{kaliakin2024accurate}
Kaliakin,~D.; Shajan,~A.; Moreno,~J.~R.; Li,~Z.; Mitra,~A.; Motta,~M.; Johnson,~C.; Saki,~A.~A.; Das,~S.; Sitdikov,~I. Accurate quantum-centric simulations of supramolecular interactions. \emph{arXiv preprint arXiv:2410.09209} \textbf{2024}, \relax
\mciteBstWouldAddEndPunctfalse
\mciteSetBstMidEndSepPunct{\mcitedefaultmidpunct}
{}{\mcitedefaultseppunct}\relax
\EndOfBibitem
\bibitem[Barroca \latin{et~al.}(2025)Barroca, Gujarati, Sharma, Ferreira, Na, Giammona, Mezzacapo, Wunsch, and Steiner]{barroca2025surface}
Barroca,~M.~A.; Gujarati,~T.; Sharma,~V.; Ferreira,~R. N.~B.; Na,~Y.-H.; Giammona,~M.; Mezzacapo,~A.; Wunsch,~B.; Steiner,~M. Surface Reaction Simulations for Battery Materials through Sample-Based Quantum Diagonalization and Local Embedding. \emph{arXiv preprint arXiv:2503.10923} \textbf{2025}, Submitted on 13 Mar 2025\relax
\mciteBstWouldAddEndPuncttrue
\mciteSetBstMidEndSepPunct{\mcitedefaultmidpunct}
{\mcitedefaultendpunct}{\mcitedefaultseppunct}\relax
\EndOfBibitem
\bibitem[Motta \latin{et~al.}(2023)Motta, Sung, Whaley, Head-Gordon, and Shee]{motta2023bridging}
Motta,~M.; Sung,~K.~J.; Whaley,~K.~B.; Head-Gordon,~M.; Shee,~J. Bridging physical intuition and hardware efficiency for correlated electronic states: the local unitary cluster Jastrow ansatz for electronic structure. \emph{Chemical Science} \textbf{2023}, \emph{14}, 11213--11227\relax
\mciteBstWouldAddEndPuncttrue
\mciteSetBstMidEndSepPunct{\mcitedefaultmidpunct}
{\mcitedefaultendpunct}{\mcitedefaultseppunct}\relax
\EndOfBibitem
\bibitem[Sun \latin{et~al.}(2018)Sun, Berkelbach, Blunt, Booth, Guo, Li, Liu, McClain, Sayfutyarova, and Sharma]{sun2018pyscf}
Sun,~Q.; Berkelbach,~T.~C.; Blunt,~N.~S.; Booth,~G.~H.; Guo,~S.; Li,~Z.; Liu,~J.; McClain,~J.~D.; Sayfutyarova,~E.~R.; Sharma,~S. PySCF: the Python-based simulations of chemistry framework. \emph{Wiley Interdisciplinary Reviews: Computational Molecular Science} \textbf{2018}, \emph{8}, e1340\relax
\mciteBstWouldAddEndPuncttrue
\mciteSetBstMidEndSepPunct{\mcitedefaultmidpunct}
{\mcitedefaultendpunct}{\mcitedefaultseppunct}\relax
\EndOfBibitem
\bibitem[Sun \latin{et~al.}(2020)Sun, Zhang, Banerjee, Bao, Barbry, Blunt, Bogdanov, Booth, Chen, and Cui]{sun2020recent}
Sun,~Q.; Zhang,~X.; Banerjee,~S.; Bao,~P.; Barbry,~M.; Blunt,~N.~S.; Bogdanov,~N.~A.; Booth,~G.~H.; Chen,~J.; Cui,~Z.-H. Recent developments in the PySCF program package. \emph{The Journal of Chemical Physics} \textbf{2020}, \emph{153}, 024109\relax
\mciteBstWouldAddEndPuncttrue
\mciteSetBstMidEndSepPunct{\mcitedefaultmidpunct}
{\mcitedefaultendpunct}{\mcitedefaultseppunct}\relax
\EndOfBibitem
\bibitem[Marenich \latin{et~al.}(2020)Marenich, Kelly, Thompson, Hawkins, Chambers, Giesen, Winget, Cramer, and Truhlar]{Manerich2020}
Marenich,~A.~V.; Kelly,~C.~P.; Thompson,~J.~D.; Hawkins,~G.~D.; Chambers,~C.~C.; Giesen,~D.~J.; Winget,~P.; Cramer,~C.~J.; Truhlar,~D.~G. Minnesota Solvation Database (MNSOL) version 2012. \textbf{2020}, Retrieved from the Data Repository for the University of Minnesota (DRUM)\relax
\mciteBstWouldAddEndPuncttrue
\mciteSetBstMidEndSepPunct{\mcitedefaultmidpunct}
{\mcitedefaultendpunct}{\mcitedefaultseppunct}\relax
\EndOfBibitem
\bibitem[{ffsims developers}(2024)]{ffsim2024}
{ffsims developers} ffsim: Faster simulations of fermionic quantum circuits. \url{https://github.com/qiskit-community/ffsim}, 2024; Accessed: 2024-09-01\relax
\mciteBstWouldAddEndPuncttrue
\mciteSetBstMidEndSepPunct{\mcitedefaultmidpunct}
{\mcitedefaultendpunct}{\mcitedefaultseppunct}\relax
\EndOfBibitem
\bibitem[Moritz \latin{et~al.}(2017)Moritz, Nishihara, Wang, Tumanov, Liaw, Liang, Elibol, Yang, Paul, and Jordan]{moritz2017ray}
Moritz,~P.; Nishihara,~R.; Wang,~S.; Tumanov,~A.; Liaw,~R.; Liang,~E.; Elibol,~M.; Yang,~Z.; Paul,~W.; Jordan,~M.~I. Ray: A Distributed Framework for Emerging {AI} Applications. \emph{arXiv preprint arXiv:1712.05889} \textbf{2017}, Available at \url{https://arxiv.org/abs/1712.05889}\relax
\mciteBstWouldAddEndPuncttrue
\mciteSetBstMidEndSepPunct{\mcitedefaultmidpunct}
{\mcitedefaultendpunct}{\mcitedefaultseppunct}\relax
\EndOfBibitem
\bibitem[Hunt(2023)]{Hunt2023}
Hunt,~J. \emph{A Beginners Guide to Python 3 Programming}; Springer International Publishing: Cham, 2023; pp 487--490\relax
\mciteBstWouldAddEndPuncttrue
\mciteSetBstMidEndSepPunct{\mcitedefaultmidpunct}
{\mcitedefaultendpunct}{\mcitedefaultseppunct}\relax
\EndOfBibitem
\bibitem[Coe(2023)]{Coe2023}
Coe,~J.~P. Analytic Gradients for Selected Configuration Interaction. \emph{Journal of Chemical Theory and Computation} \textbf{2023}, \emph{19}, 874--886, PMID: 36656261\relax
\mciteBstWouldAddEndPuncttrue
\mciteSetBstMidEndSepPunct{\mcitedefaultmidpunct}
{\mcitedefaultendpunct}{\mcitedefaultseppunct}\relax
\EndOfBibitem
\bibitem[Sun(2015)]{Libcint15}
Sun,~Q. Libcint: An efficient general integral library for Gaussian basis functions. \emph{Journal of Computational Chemistry} \textbf{2015}, \emph{36}, 1664--1671\relax
\mciteBstWouldAddEndPuncttrue
\mciteSetBstMidEndSepPunct{\mcitedefaultmidpunct}
{\mcitedefaultendpunct}{\mcitedefaultseppunct}\relax
\EndOfBibitem
\end{mcitethebibliography}

\newpage
\begin{center}
\section*{Supplementary Information: Alchemical Free Energy Calculations Using Quantum Hardware}
\end{center}

\subsection{QUICK interface with PySCF and Qiskit Addon: SQD}

\begin{figure}
    \centering
    \includegraphics[width=1\linewidth]{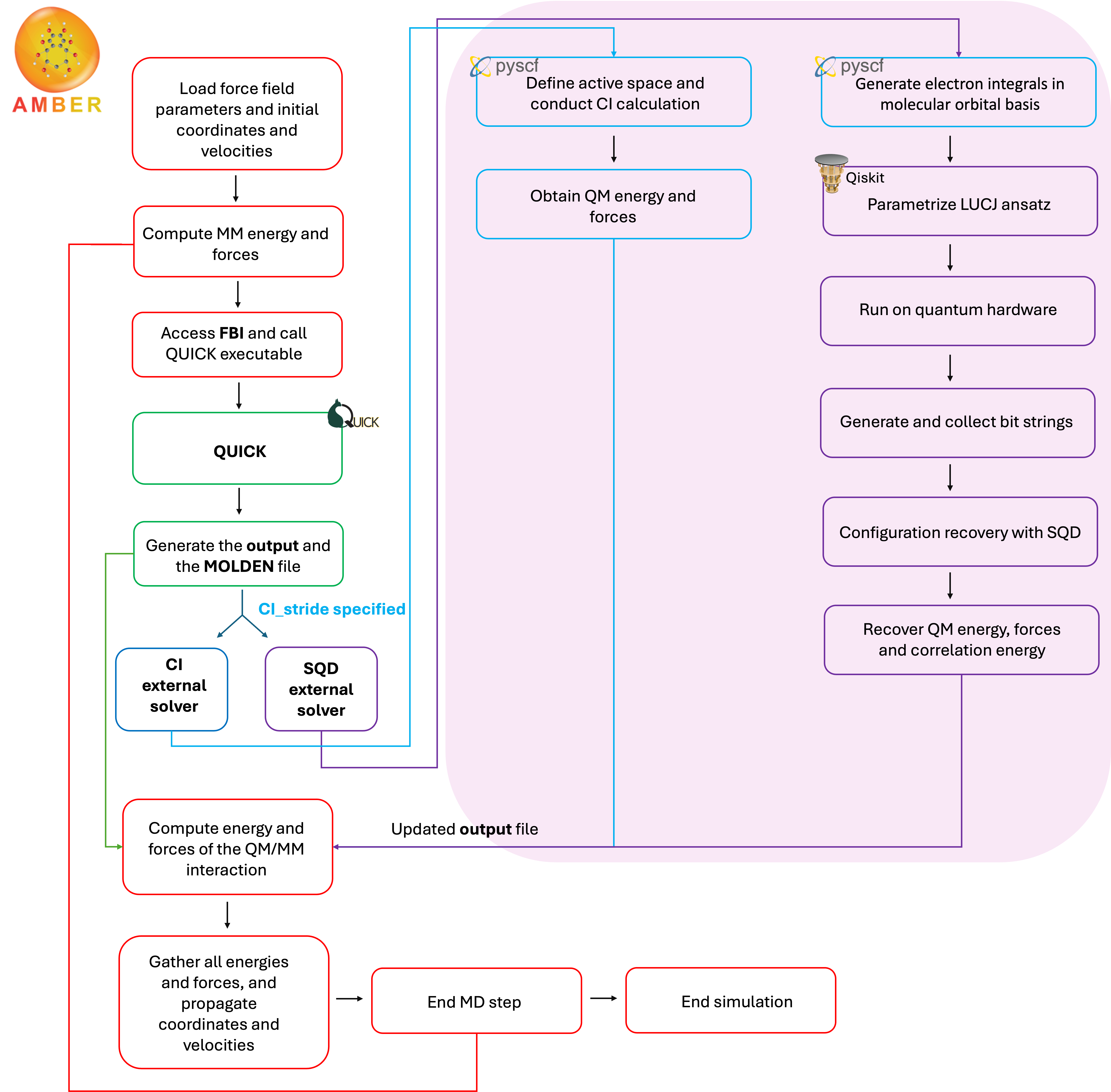}
    \caption*{Figure S1: Schematic representation of the integration of AMBER’s sander module, the QUICK quantum engine, and external CI solvers. The CI\_stride keyword controls the redirection of the QM computation to one of the two external CI solvers. The color scheme highlights the workflow: steps managed by sander are shown in red, those performed by QUICK engine in green. The purple box represents the steps performed by either the full FCI calculations, or by quantum-centric SQD workflow. PySCF code is reported in light blue, whereas the steps controlled by Qiskit are in purple.}
    \label{figS1}
\end{figure}

To enable coupling between  MD simulations performed on classical hardware and quantum circuit-based computations, we developed a custom interface that connects AMBER/QUICK with Qiskit \cite{aleksandrowicz2019qiskit,javadi2024quantum}. In addition, we integrated simulations with PySCF 2.8.0 \cite{sun2018pyscf,sun2020recent} into AMBER/QUICK workflow to enable FCI calculations (Figure S1). The standard file-based interface (FBI) between AMBER \cite{amber24} and QUICK \cite{manathunga2023quantum,cruzeiro2021open} was extended to include both two new modules: A) the FCI solver, implemented through the interface with PySCF, and B) the SQD solver, that utilizes the quantum hardware to generate the initial electron configurations with LUCJ Ansatz and subsequent post-processing through Qiskit Addon: SQD. 

The AMBER Fortran module, specifically qm2\_extern\_quick\_module.F90, was modified to interface the classical engine (AMBER/QUICK) with the PySCF code, managing the exchange of data between the classical simulation and the quantum backends. A dedicated keyword CI\_stride was specified within AMBER to define the frequency (in MD steps) at which the CI or SQD solver is triggered. QUICK's FBI mode supports custom input template to define the QM method, basis set, and target measures to compute such as energy and gradient. This template also allows the generation of a MOLDEN file, which contains the molecular orbital and geometry data essential for both the CI and SQD solver. 
Shared functionalities are implemented in the quick\_parsing\_utilities.py script. This script reads the QUICK\_job.out file, extracts the coordinates and any associated MM point charges, forwards them to the external solver, and finally integrates the computed energy, nuclear gradient and correlation energy into a new output file, QUICK\_job\_PySCF\_modified.out. Depending on the simulation environment (either in aqueous solution or in vacuum) the system is initialized differently. In the gas-phase simulations, no external charges are present, and a standard RHF is performed. When water molecules are present, their coordinates and charges are passed to the molecular system definition within PySCF and the system is initialized using PySCF QM/MM module, enabling electrostatic embedding of the classical environment. Once the HF object is constructed, it is populated with orbital coefficients, energies, and occupations extracted from the MOLDEN file. In the FCI solver, a FCI calculation is performed using  CI\_solver\_PySCF.py and the result is integrated into AMBER through QUICK\_job\_PySCF\_modified.out influencing the subsequent MD propagation. For the SQD solver, the quantum-centric simulation, two Python scripts are provided. LUCJ-run.py prepares the quantum input by saving the one-electron and two-electron integrals of the orbitals within the active space. These integrals are written in a FCIDUMP file together with the nuclear repulsion and the electron number. In the following step the FCIDUMP is read back to run a Coupled Cluster Singles and Doubles (CCSD) calculation, from which the single t\textsubscript{1} and double t\textsubscript{2} excitation amplitudes are extracted. 
We generate the LUCJ circuits using the ffsim library (version 0.0.49) \cite{ffsim2024} interfaced with Qiskit 1.3.3 \cite{aleksandrowicz2019qiskit,javadi2024quantum}. The t\textsubscript{2} amplitudes are used to parametrize the LUCJ ansatz, constructed via the \texttt{UCJOpSpinBalanced} class. The circuits include 2 repetitions (\texttt{n\_reps=2}) and entangling connections defined by nearest-neighbor and on-site interaction patterns in the active space. In the present work we used \texttt{ibm\_strasbourg} quantum processor. Circuit transpilation is performed with optimization level 3 and includes a custom pre-initialization step provided by ffsim. Quantum error mitigation is applied through gate twirling (while measurement twirling is disabled), as enabled by the \texttt{SamplerV2} primitive in Qiskit's runtime library (version 0.36.1). The final circuit layout is reported in Figure S2 and more circuit details are described in Figure S3.

\begin{figure}
    \centering
    \includegraphics[width=1\linewidth]{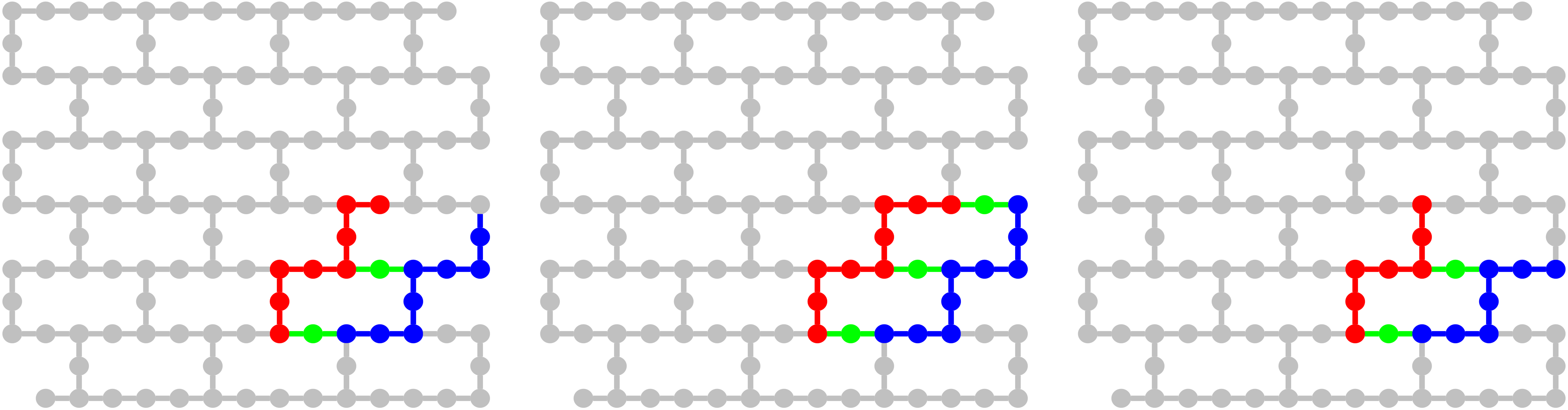}
    \caption*{Figure S2:
Qubits layout of the LUCJ circuit for each of three solutes investigated in the present study. A) (10e,8o) ammonia, B) (10e,9o) methane  and, C) (10e,7o) water. Qubits corresponding to the occupation numbers of $\alpha$ and $\beta$ spin-orbitals are represented in red and blue, respectively. Auxiliary qubits employed to mediate the density-density interactions between $\alpha$ and $\beta$ spin-orbitals are depicted in green.}
\renewcommand{\figure}{S2}

    \label{figS2}
\end{figure}

\begin{figure}
    \centering
    \includegraphics[width=0.5\linewidth]{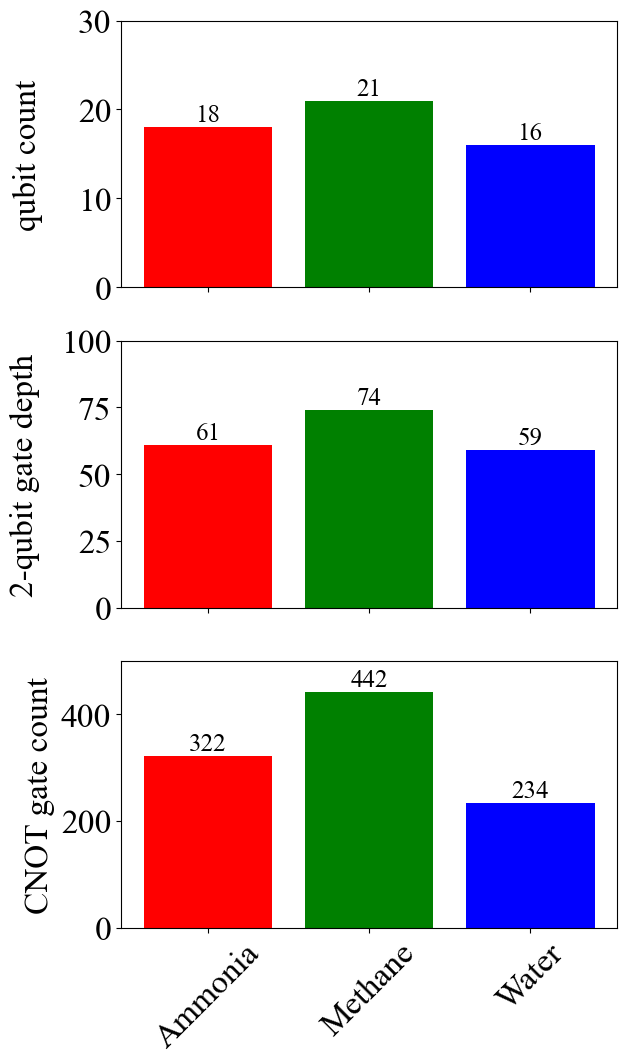}
    \caption*{Figure S3: Qubits number, 2-qubits gate depth and CNOT gate count for the LUCJ circuit of (10e,8o) ammonia in red, (10e,9o) methane in green, and (10e,7o) water in blue.}
    \label{figS3}
\end{figure}
The post-processing is handled by the \texttt{CI\_solver\_SQD.py} script, which uses qiskit-addon-sqd 0.9.0 library  and reads the \texttt{count\_dict.txt}. The output is a collection of quantum states generated by the run on the quantum hardware. These bitstrings include both physical and non-physical configurations due to noise inherent in current quantum hardware. To refine the data, the script first retrieves only the bitstrings with correct particle number forming initial orbital occupations. The subsequent steps exploit prior orbital occupations to perform self-consistent configuration recovery (S-CORE) procedure which gradually recovers the correct spin-up and spin-down electron counts in the noise-corrupted bitstrings based on the distance from the current value of the given bit to the average occupancy of the spin-orbital.\cite{robledo2024chemistry} From this cleaned set, batches of configurations defining the subspaces of the full Hilbert space are generated. For each subspace, the Hamiltonian is projected and diagonalized to obtain the ground state. These ground states are then analyzed to extract average orbital occupations, which guide the probabilistic regeneration of a new, refined ensemble of bitstrings. The cycle of subspace construction, Hamiltonian projection, diagonalization, occupation averaging, and configuration recovery is repeated iteratively until a predefined convergence criterion is met, progressively reducing noise and improving the a of the quantum simulation results. In the present work we used 10 batches and 100 samples per batch.  We parallelized the calculations across 10 CPUs using Ray~\cite{moritz2017ray}, performing two iterations. In this work, we introduced a monkey patch \cite{Hunt2023} to use customized functions within PySCF code. Specifically, we modified the CI-wrapper of PySCF and integrated \texttt{kernel\_fixed\_space} inside of it, which allowed us to access the gradient module of PySCF.\cite{sun2018pyscf,sun2020recent} The latter supports analytical gradients for SCI calculations \cite{Coe2023,Libcint15}, and we were able to extend it to SQD calculations as well. The gradient is calculated and integrated into the MD simulation only for the last iteration of S-CORE within SQD, since it corresponds to the optimal subspace that underwent the configuration recovery. More details regarding the number of samples and configurations are reported in Table S1. In present paper the simulations performed with all of the orbitals within sto-3g basis set. However, both PySCF \cite{sun2018pyscf,sun2020recent} CI and SQD \cite{robledo2024chemistry, kaliakin2024accurate,yu2025quantum,Liepuoniute2024,Barison2025,barroca2025surface} solvers support an "all-orbitals" and a "frozen core" configuration, the latter freezes the 1s core orbital of heavy atoms to reduce computational cost while preserving accuracy with larger basis sets.

\begin{table}[t]
\centering
\caption*{Table S1: Details of SQD calculations. For each system, we report the Active Space (AS); $D_{\text{AS}}$ represents the Hilbert space dimension. $|\widetilde{\chi}_b|$ is the number of samples per batch. $\mathit{d}$ represents the subspace (batch) with the lowest energy across all batches at the last iteration. The reported values for $\mathit{d}$  correspond to the first step of the production run in aqueous environment.}
\label{tableS1}
\begin{tabular}{lccc c}
\hline\hline
\textbf{System} & \textbf{AS} & $\boldsymbol{|\widetilde{\chi}_b|}$ & $\boldsymbol{\mathit{d}}$ & $\boldsymbol{D_{\text{AS}}}$ \\
\hline
Ammonia & $(10e, 8o)$ & $100$ & $2500$& $3136$ \\
Methane & $(10e, 9o)$ & $100$ & $7056$& $15876$ \\
Water   & $(10e, 7o)$ & $100$ & $441$& $441$ \\
\hline\hline
\end{tabular}

\end{table}

\newpage
\subsection{Quantum-centric SQD simulations }

At the heart of SQD \cite{robledo2024chemistry, kaliakin2024accurate,yu2025quantum,Liepuoniute2024,Barison2025,barroca2025surface,yu2025quantum} lies the execution of a quantum circuit designed to sample a set of  states $\chi = \{ \bts_1 \dots \bts_d \}$  that serve as the computational basis for the diagonalization of the molecular Hamiltonian. 
An initial wavefuntion guess $ \left| \Phi_{qc} \right\rangle $, approximating of the ground state,  is prepared from a truncated version of the local unitary cluster Jastrow (LUCJ) ansatz. Here the mapping of fermions to qubits is performed with the standard Jordan-Wigner (JW) mapping and the LUCJ ansatz \cite{motta2023bridging} has a following form:
\begin{align}
|\Phi_{\mathrm{qc}}\rangle = e^{-\hat{K}_2} e^{\hat{K}_1} e^{i\hat{J}_1} e^{-\hat{K}_1} |x_{\mathrm{RHF}}\rangle,
\end{align}
where the one-body operators are denoted by $ \hat{K}_1 $ and $ \hat{K}_2$ , while $ \hat{J}_1 $ represents the density-density operator, and $|x_{\mathrm{RHF}}\rangle$ is the restricted closed-shell RHF state.  The LUCJ ansatz was parametrized using amplitudes obtained from the gas-phase restricted closed-shell CCSD calculations performed within selected active space, following the approach adopted in prior quantum-centric SQD studies \cite{robledo2024chemistry, kaliakin2024accurate,yu2025quantum,Liepuoniute2024,Barison2025,barroca2025surface,yu2025quantum}. We run the quantum circuit on a quantum computer and measure the state \( |\Psi\rangle \) in the computational basis. By repeating the measurement many times, we collect a set of bitstrings
\[
\tilde{\chi} = \{ \mathbf{x} \mid \mathbf{x} \sim \tilde{P}_\Psi(\mathbf{x}) \},
\]

where each bitstring \( \mathbf{x} \in \{0,1\}^M \) represents an electronic configuration (Slater determinant) sampled according to \( \tilde{P}_\Psi \). However, the results obtained from the execution of a quantum circuit are inherently affected by errors arising from noise present in current quantum devices. This noise can introduce broken particle-number and spin-z symmetries in the samples, spreading the distribution  \( {P}_\Psi \) over configurations that do not participate to the low-energy states.  As a consequence, only a fraction of the $\widetilde{\chi}$ contribute to the ground state. To mitigate this, we applied a configuration recovery protocol that iteratively recovers configurations with the correct particle number. For each configuration \( \mathbf{x} \in \tilde{\chi} \) such that \( N_{\mathbf{x}} \neq N \), a number \( |N_{\mathbf{x}} - N| \) of spin-orbitals are flipped. Bit-flip probabilities are weighted according to a monotonically increasing function of the difference $|x_{p\sigma} - n_{p\sigma}|$ where  $x_{p\sigma}$  is the bit value and  $n_{p\sigma}$  is the average occupation of spin-orbital $p\sigma$  from the previous recovery round. The initial guess of the occupancies n used in the first recovery iteration is computed from the raw quantum samples  $\widetilde{\chi}$  \cite{robledo2024chemistry}. From this process we obtain the refined configurations $\chi_R $ form which we build K batches of d configurations, $S(1), \ldots, S(K)$, according to a distribution proportional to the empirical frequencies of each x in $\chi_R $.  Each batch spans a many-body subspace over which the Hamiltonian is projected and diagonalized:

\begin{equation}
\hat{H}_{S^{(b)}} = \hat{P}_{S^{(b)}} \hat{H} \hat{P}_{S^{(b)}},
\end{equation}

Where the projector $\hat{P}_{S^{(b)}}$ is,
\begin{equation}
\hat{P}_{S^{(b)}} = \sum_{x \in S^{(b)}} |x\rangle \langle x|.
\end{equation}

The diagonalization of  $\hat{H}_{S^{(b)}}$ yields the subspace ground state $|\psi^{(b)}\rangle$ and corresponding energy $E^{(b)}$, computed using the iterative Davidson method. 

We use the lowest energy across batches as $min_{b}$  $E^{(b)}$, as the best approximation to the ground state energy. The wavefunctions $|\psi^{(b)}\rangle$ from all the batches are then used to get the new occupancies,
\begin{equation}
n_{p\sigma} = \frac{1}{K} \sum_{b=1}^{K} \langle \psi^{(b)} | \hat{n}_{p\sigma} | \psi^{(b)} \rangle
\end{equation}

These occupancies are fed back into the configuration recovery step. The self-consistent procedure is repeated until reached convergence. 
\\
\\
\\
\subsection{Book-end corrections values}
\begin{table}[h!]
\centering
\caption*{Table S2: Book-ending correction values computed with three different protocols. All values are expressed in kcal/mol.}
\label{tableS1}
\begin{tabular}{lccc}
\hline\hline
\textbf{System}& \makecell{\textbf{HF} \\ \textbf{protocol}} & \makecell{\textbf{HF + FCI} \\ \textbf{protocol}} & \makecell{\textbf{HF + SQD} \\ \textbf{protocol}} \\
\hline
Ammonia & $1.93 \pm 0.30$& $1.52 \pm 0.36$& $1.67 \pm 0.33$ \\
Methane & -$0.15 \pm 0.07$& -$0.84 \pm 0.33$& -$0.37 \pm 0.90$\\
Water   & $1.94 \pm 0.30$& $1.38 \pm 0.13$& $1.52 \pm 0.07$ \\
\hline\hline
\end{tabular}
\end{table}

\end{document}